\documentclass[twocolumn,twocolappendix]{aastex631}

\usepackage{graphicx}
\usepackage{hyperref}

\newcommand\gdrthree{Gaia~DR3}
\newcommand\MH{\ensuremath{\mathrm{[M/H]}}}
\newcommand\logg{\ensuremath{\log g}}
\newcommand\Teff{\ensuremath{T_\textrm{eff}}}
\received{February 3, 2023} 
\accepted{May 11, 2023}

\submitjournal{ApJS}

\begin{document}


\title{Robust Data-driven Metallicities for 175 Million Stars from Gaia XP Spectra}

\correspondingauthor{Ren\'e Andrae}
\email{andrae@mpia.de}

\author[0000-0001-8006-6365]{Ren\'e Andrae}
\affiliation{Max-Planck-Institut f{\"u}r Astronomie, K{\"o}nigstuhl 17, D-69117 Heidelberg, Germany}

\author[0000-0003-4996-9069]{Hans-Walter Rix}
\affiliation{Max-Planck-Institut f{\"u}r Astronomie, K{\"o}nigstuhl 17, D-69117 Heidelberg, Germany}

\author[0000-0002-0572-8012]{Vedant Chandra}
\affiliation{Center for Astrophysics $\mid$ Harvard \& Smithsonian, 60 Garden St, Cambridge, MA 02138, USA}

\begin{abstract}
We derive and publish data-driven estimates of stellar metallicities \MH\ for $\sim 175$ million stars with low-resolution XP spectra published in Gaia DR3. The \MH\ values, along with \Teff\ and \logg,  
are derived using the XGBoost algorithm, trained on stellar parameters from APOGEE, augmented by a set of very metal-poor stars. 
XGBoost draws on a number of data features: the full set of XP spectral coefficients, narrowband fluxes derived from XP spectra, and broadband magnitudes.  In particular, we include CatWISE magnitudes, as they reduce the degeneracy of \Teff\ and dust reddening.  We also include the parallax as a data feature, which helps constrain \logg\ and \MH. The resulting mean stellar parameter precision is 0.1\,dex in \MH, 50\,K in \Teff, and 0.08\,dex in \logg.
This all-sky \MH\ sample is substantially larger than published samples of comparable fidelity across $-3\lesssim \MH \lesssim +0.5$. 
Additionally, we provide a catalog of over 17 million bright ($G<16$) red giants whose \MH\ are vetted to be precise and pure.
We present all-sky maps of the Milky Way in different \MH\ regimes that illustrate the purity of the dataset, and demonstrate the power of this unprecedented sample to reveal the Milky Way's structure from its heart to its disk.
\end{abstract}



\section{Introduction} \label{sec:intro}

The chemical composition of stars, reflected in their photospheric abundances, is a fundamental stellar observable. To zeroth order, it can be summarized by the mean \emph{metallicity} \MH, which varies by orders of magnitude among the stellar populations in the Galaxy, while the individual abundance ratios among heavy elements tend to vary far less.

In the context of the Milky Way, or other resolved nearby galaxies such as the Magellanic Clouds, having vast samples of stars with \MH\ estimates across all stellar populations matters greatly for both galaxy and stellar evolution: \MH\ traces the ``chemical evolution'' of the galaxy that reflects the combination of the star formation history, stellar yields, gas inflow and feedback \citep[e.g,][]{Dekel1986, Matteucci1994, Tremonti2004}. Large and systematically selected samples of low-\MH\ stars are needed to quantify and test stellar yields and the importance of different nucleosynthetic channels, in particular of the earliest stars in a galaxy \citep[e.g.,][]{Tinsley1979,McWilliam1997}. Furthermore, \MH\ is indispensable to study the \emph{chemo-dynamics} of, say, the Milky Way: the determination and evolutionary interpretation of the stars' distribution in the space of orbits, ages, and element abundances \citep[e.g.,][]{Hayden2015,Weinberg2019,Weinberg2022}. Studies of the chemical and dynamical evolution of the Galaxy are linked closely, as the abundances, in particular \MH, also serves as an -- albeit complex -- proxy for stellar ages \citep[e.g.,][]{Tinsley1980,Twarog1980,Nordstrom2004,Gallazzi2005,Rix2022}. 

While \MH\ is the fundamental measurement for abundances, the $\alpha$-element enhancement has long been established as the arguably next most important abundance observation, as it reflects the relative roles of core-collapse and thermonuclear supernovae in the enrichment of a star's birth material \citep{Tinsley1979,Hayden2015,Weinberg2019}.  The dimensionality of the abundance space for elements through the iron peak is not yet fully settled \citep{Ness2019,Ting2022}. The abundances of elements beyond the iron peak, which arose primarily via the $s$- and $r$-process, is of great interest \citep[e.g.,][]{Sneden2008}. However, the observational determination of these elements requires spectra of relatively high resolution and S/N \citep[e.g.,][]{Ji2019a, Ji2019b}.

These arguments have motivated over the last decade(s) a suite of large-scale spectroscopic surveys: SDSS I-IV \citep{York2000}, LAMOST \citep{Cui2012}, GALAH \citep{DeSilva2015}, Gaia-ESO \citep{Gilmore2012} in the past; SDSS-V \citep{SDSSV}, WEAVE \citep{Dalton2012} now; 4MOST \citep{deJong2019} in the future.
Although these ground-based surveys have now reached sample sizes of nearly $10^7$ stars, all have had highly incomplete sky coverage and complex selection functions. Only SDSS-V will provide ground-based spectroscopic all-sky coverage over the next few years \citep{SDSSV,SDSS_DR18}. 

As Gaia's most recent data release DR3 \citep{GaiaDR3} has again made abundantly clear, Gaia is not only a photometric and astrometric mission, but also a spectroscopic one. Gaia obtained spectra both with the RVS instrument, at a resolution of $\sim$8000 around the near-IR Ca triplet \citep[Seabroke et al.\ in prep.,][]{2022gdr3.reptE...6S}; 
and very low-resolution spectra ($R\sim 40-150$) taken with the two prisms BP and RP \citep{2021A&A...652A..86C,2022arXiv220606143D} that together cover the wavelength range from $\sim$350nm to $\sim$1,000nm \citep{Montegriffo2022}. In the following, we denote these BP and RP data as \emph{XP} spectra.
These spectra --- and astrophysical parameters derived from these spectra --- were released in Gaia DR3 for both RVS and XP: 220 million and 1 million spectra, as well as 470 million and 6 million sets of astrophysical parameters, for XP and RVS, respectively.

To maximize the Gaia data set suitable for chemo-dynamical studies of our Galaxy, one would need to have abundances, at least \MH, for most stars that have RVS velocities\footnote{Essentially all stars with RVS velocities also have parallax and proper motions, needed to calculate orbits along the the RVS velocity.}. In the context of DR3, this can be done with \MH\ based on XP spectra, but not on RVS spectra or derived abundances, as these are published only for a subset of 6 million stars \citep{Recio-Blanco-GSPspec} compared to 33.8 million stars that have RVS velocities \citep{2022arXiv220605902K}.

An extensive set of metallicities for Gaia sources with XP spectra was published as part of DR3 \citep{Andrae22}. By design, these \MH\ values were derived using synthetic model spectra in comparison with the XP specta, with the goal of a consistent approach to stellar parameter estimates across much of the CMD. Unfortunately, external validation has shown that these \MH\ values have important shortcomings (e.g.\ systematics and a high rate of ``catastrophic'' outliers) due to two aspects known and stated at the time of publication: First, knowledge of the Gaia XP system is detailed but imperfect, so that significant discrepancies between the predictions of the synthetic model and the XP data exist and lead to erroneous \MH\ estimates. Second, the spectra of different \Teff\ and \logg\ have very different information content about \MH\ at low resolution and for some temperatures (e.g.\ OB stars), the XP spectra are simply not informative on \MH.

On the other hand, it has been established that for cool stars, even very low-resolution spectra are informative about \MH\ \citep{Ting2017}. By using a data-driven approach to estimate \MH\ and by focusing on stellar types whose low-resolution spectra are informative about \MH, one can overcome these limitations. This has recently been shown by \cite{Rix2022}, who produced a large set of \MH\ estimates towards the Galactic center. Similar methods have been employed to study the halo of the Galaxy, for example to map out its last major merger \citep{Belokurov2023,Chandra2023}.

Here we set out to build on this work and produce a comprehensive catalog of high-fidelity stellar \MH\ estimates (with \Teff\ and \logg, as corollary) that 
\begin{itemize}
    \item includes essentially the entire sample of published XP spectra, acknowledging that the \MH\ estimates at low S/N and high \Teff\ are potentially unreliable. This requires the identification of subsamples where the \MH\ estimates are precise, accurate and robust (i.e.\ with negligible outliers), as verified by external comparison.
    \item is ``all-sky'', accepting that the reach of such a catalog varies across the sky due to a) the Gaia experimental set-up; b) the details of the DR3 data release; c) the changing source density and dust extinction.
    \item is data-driven, drawing on high-quality training sets that cover essentially the whole metallicity range present in the local group $-3<\MH <0.5$.
    \item draw mostly on XP spectral information, but also utilize relevant information that is available for most of the target sample: broad-band photometry across a wide range of wavelengths (extending to WISE in the infrared) to constrain the overall SED, and reduce \Teff\ -- reddening degeneracy; the parallaxes $\varpi$ which are -- even at low or negative $\varpi/\delta\varpi$ highly informative about the luminosity or absolute magnitude $M$ of the star, as $10^{M_\lambda/5}\propto \varpi~10^{m_\lambda/5}$. Indirectly, $\varpi$ therefore informs \logg\ and \MH.
\end{itemize}
Since the first submission of this manuscript, two similar works have been published:
\citet{2023arXiv230317676Y} classify 188\,000 candidates for very-metal-poor stars ($\textrm{[Fe/H]}<-2$) using XP spectra and XGBoost while \citet{2023arXiv230303420Z} build an empirical forward model from LAMOST training examples in order to estimate stellar parameters with realistic uncertainties for all 220 million published XP spectra.

The rest of the paper is organized as follows: In Sect.~\ref{sec:training-xgboost}, we explain how we compile the training sample, what input features we choose for XGBoost and we show first internal validation results. In Sect.~\ref{sec:application}, we define our application sample and validate our results on external data that have not been used for testing. In the closing Sect.~\ref{sec:illustration}, we illustrate the power of this sample by showing a set of all-sky maps
in different metallicity bins, which illustrates that even the (rare) low-metallicity subsamples have little if any contamination. In the summary and outlook, we touch on obvious future astrophysical uses of this sample. The catalogs produced in this work are published online\footnote{\url{https://doi.org/10.5281/zenodo.7599788}} \citep{Andrae2023Zenodo}.

\section{Training XGBoost}
\label{sec:training-xgboost}

We seek to train XGBoost models \citep{Chen:2016:XST:2939672.2939785} to estimate stellar metallicity, effective temperature and surface gravity from XP spectra, drawing on the subset of objects for which both XP spectra and  externally-derived stellar parameters of high fidelity exist. 

\subsection{Training sample selection}
\label{ssec:training-sample}

Encouraged from the results in \citet{Rix2022}, we train XGBoost models using as data features both XP coefficients \citep{2021A&A...652A..86C,2022arXiv220606143D} and synthesized photometry that was computed with GaiaXPy \citep[e.g.][]{2022arXiv220606215G}. For the most part, we do this for stars with literature labels from APOGEE DR17 \citep{SDSS_DR17}. However, APOGEE DR17 has no metallicity estimates below $\MH\sim-2.5$, and XGBoost cannot extrapolate beyond the metallicity range of its training sample,\footnote{XGBoost is a tree-based method. As such, it segments the feature space and assigns as output to each segment the average training labels within that segment.}. Therefore, we augment the APOGEE DR17 training sample by the set of very/ultra-metal-poor stars from \citet{2022ApJ...931..147L}, which provide a consistent and fairly extensive set of \MH\  determinations for a set of (apparently) bright stars. We also replace the AllWISE photometry \citep{2014yCat.2328....0C} using in \citet{Rix2022} by CatWISE photometry \citep{2021ApJS..253....8M}, which is deeper and thus achieves higher completeness.

APOGEE DR17 contains a total of 733\,901 stars. Of these, 647\,025 actually have stellar parameters \Teff, \logg, and \MH, and 643\,401 have a cross-match to Gaia. Among these, only 599\,662 achieve signal-to-noise ratios above 50 in the APOGEE spectra. For the purpose of this paper, we also require XP spectra to be available, which is the case in the Gaia DR3 data for 537\,412 of these stars. We also require CatWISE photometry in $W_1$ and $W_2$ bands as these bands greatly aid reducing the temperature -- extinction degeneracy. This reduces the set of APOGEE DR17 training to 510\,413, which only contains 485\,850 unique Gaia source IDs, i.e.\ there are ``duplicates'' which represent repeated APOGEE observations of the same star. In such cases, we adopt the mean APOGEE parameters averaged over all repeat observations as training labels.
The results of \citet{2022ApJ...931..147L} encompass 385 stars, of which 291 stars have published XP spectra, as well as $W_1$ and $W_2$ photometry in CatWISE.

\begin{figure}
\begin{center}
\includegraphics[width=\columnwidth]{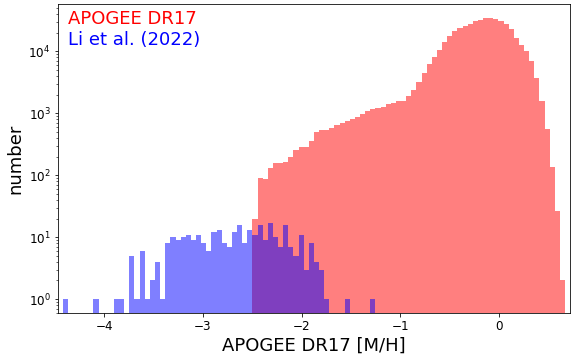}
\end{center}
\caption{Distribution of \MH\ in training sample of stars, which is drawn from SDSS-APOGEE DR17 (red) and the very metal-poor stars from \citet{2022ApJ...931..147L} (blue).}
\label{fig:distribution-MH-training-sample}
\end{figure}

The resulting \MH\ distribution of this training sample is shown in Fig.~\ref{fig:distribution-MH-training-sample}. Already from APOGEE, we have good coverage for $\MH < -1$. But this figure also shows how critical the inclusion of stars from \citet{2022ApJ...931..147L} is to cover metallicities below $-2.5$, eventually down to a minimum value of $-4.37$. This expanded training sample removes an important limitation at low metallicity of the work by \citet{Rix2022}.

\begin{figure*}
\begin{center}
\includegraphics[width=2\columnwidth]{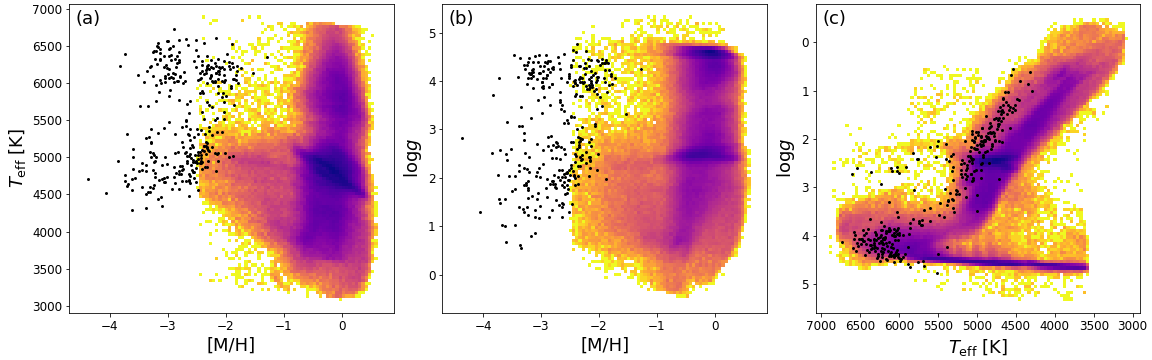}
\end{center}
\caption{Distributions of the \MH\ training sample terms of effective temperature, surface gravity and metallicity. The dominant SDSS-APOGEE DR17 part of the sample is shown as the logarithmic density map; the metal-poor training stars from \citet{2022ApJ...931..147L} as black dots.}
\label{fig:training-sample-Teff-logg-MH}
\end{figure*}

It is worth taking a closer look at the distribution of all stellar parameters in our training sample in Fig.~\ref{fig:training-sample-Teff-logg-MH}, to understand over which range we can expect XGBoost to return robust estimates. While Fig.~\ref{fig:training-sample-Teff-logg-MH}c suggests that we have a good coverage of main-sequence dwarfs and red-giant stars, the temperature range is limited to 3107K to 6867K. In particular, we have no OBA stars, no white dwarfs, or ultra-cool dwarfs in our training sample. Furthermore, Fig.~\ref{fig:training-sample-Teff-logg-MH}a shows that we have essentially no training examples for $\MH<-2$ and $\Teff<4000$K. Also, Fig.~\ref{fig:training-sample-Teff-logg-MH}b shows that we have only very few training examples for metal-poor dwarfs with $\logg>3.5$ and $\MH<-1$. This will likely preclude robust and precise parameter estimates in this regime. 

\subsection{Completeness of photometry}
\label{ssec:completeness-photometry}

\begin{figure}
\begin{center}
\includegraphics[width=\columnwidth]{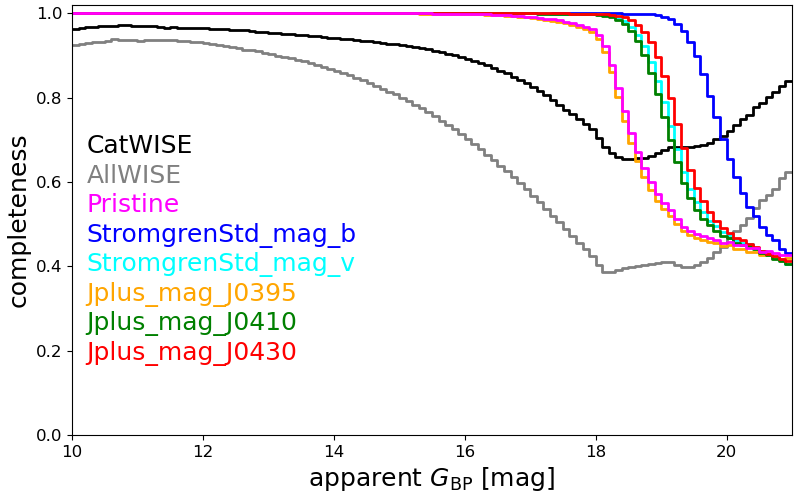}
\end{center}
\caption{Completeness of the sample at a given $G_\textrm{BP}$ magnitude in several bandpasses that limit the application of XGBoost,
which requires the \emph{full} set of features for both training and testing. For $G_\textrm{BP}\lesssim 17.7$ CatWISE is the most severe limitation, while for $G_\textrm{BP}\gtrsim 17.7$ the two narrow bandpasses synthesized with GaiaXPy in the far blue (e.g.\ \texttt{Pristine\_mag\_CaHK} and \texttt{Jplus\_mag\_J0395}) become severly incomplete.}
\label{fig:completeness_vs_GBP}
\end{figure}

\citet{Rix2022} noticed that some bands synthesized from the XP coefficients with GaiaXPy had negative fluxes and thus invalid magnitudes, particularly narrow bands in the blue where fluxes are often low. This leads to a rapidly decreasing completeness of XGBoost predictions in \citet{Rix2022} at the faint end.  Here, we address this issue more systematically to keep the completeness towards the faint end as high as possible. First, for all stars in our application sample, we synthesize their photometry with GaiaXPy in the following photometric systems that we {\it a priori} believe to be useful for estimating metallicities \citep[for details see][]{2022arXiv220606215G}:
\begin{itemize}
\item Pristine
\item Stromgren\_Std 
\item JPLUS
\item PanSTARRS1\_Std
\item Sky\_Mapper
\item Gaia\_2
\end{itemize}
Second, for all photometric bands, derived from the XP spectra, CatWISE and AllWISE, we investigate the completeness of its magnitudes as function of $G_\textrm{BP}$ in Fig.~\ref{fig:completeness_vs_GBP}.  Evidently, the completeness of synthesized photometry diminishes much earlier in some bands than in others. Further investigation reveals that bands with pivot wavelengths below $\approx$420nm are the first to be affected by incompleteness. This is consistent with our interpretation of the incompleteness arising through noise in the XP coefficients, given that the BP spectrum has low transmission and thus low signal-to-noise for wavelengths below 420nm. Furthermore, Fig.~\ref{fig:completeness_vs_GBP} shows that AllWISE would limit the completeness at all magnitudes, and that CatWISE can reach much higher completeness especially at the faint end where there are numerous stars. Still, even CatWISE does not reach full completeness even at the bright end.

\subsection{Input features for XGBoost}

For the final set of input features for XGBoost, we adopt all bands that achieve a completeness of 95\% or higher at $G_\textrm{BP}=18$. These are 31 bands and for each band, the XGBoost input feature is the color obtained from apparent $G$ magnitude minus the magnitude in this band. We choose the $G$ magnitude for all colors for two reasons: first, the $G$ magnitude is measured independently from the XP spectra from which all synthetic photometry is derived; and second, the $G$ magnitudes have very high S/N. Additionally, we use the three Gaia colors $G-G_\textrm{BP}$, $G-G_\textrm{RP}$ and $G_\textrm{BP}-G_\textrm{RP}$ as input features, as well as several colors including CatWISE photometry (namely $W_1-W_2$, $G-W_1$, $G-W_2$, $G_\textrm{BP}-W_2$). Therefore, our final set of data features comprises 38 colors and all 110 XP coefficients normalized to $G=15$. This may appear confusing at first, because the synthesised photometry is fully redundant with the XP coefficients and adding redundant features could even be detrimental to the scientific performance (curse of dimensionality). Ultimately though, the choice to include both, XP coefficients and photometry synthesised from XP, as input features for XGBoost is a matter of feature selection that we test during cross-validation \citep[see Table~1 in][]{Rix2022}. As it turns out, both are required in order to achieve optimal \MH\ results and the omission of either XP coefficients or synthetic photometry would lead to a noteworthy increase in the \MH\ errors during cross-validation and later application. This implies that XGBoost is unable to fully extract all information from the XP coefficients alone. Instead, our manual help to ``re-phrase'' the information in terms of synthesized photometry is required in order to make the information more easily accessible for XGBoost.

For deriving stellar parameters, in particular \logg , the absolute magnitude is highly informative: e.g.\ it straightforwardly differentiates between giants and dwarfs. While a substantive subset of the stars with XP spectra have good parallax S/N, form which we can estimate absolute magnitudes, many sample members have parallaxes that are consistent with zero or even negative. Therefore, we added a data feature that reflects or places limits on the absolute magnitude, but is \emph{linear} in the parallax in order to remain well behaved in cases of noisy or even negative parallaxes. Specifically, we opted for input features of the form
\begin{equation}\label{eq:feature-with-parallax}
\varpi\cdot 10^{m_X/5} 
= 10^{\log_{10}\varpi + m_X/5}
= 10^{(M_X+A_X-5)/5}
\end{equation}
where $\varpi$ denotes the parallax, $m_X$ is the apparent magnitude in some band $X$, while $M_X$ and $A_X$ are the absolute magnitude and dust attenuation in the same band $X$.\footnote{The absolute magnitude and extinction in Eq.~(\ref{eq:feature-with-parallax}) remain unknown. We merely highlight the astrophysical meaning of these features.} We added five such input features, for the photometric bands $X=G, G_\textrm{BP}, G_\textrm{RP}, W_1, W_2$. The parallax in Eq.~(\ref{eq:feature-with-parallax}) has been corrected for the parallax zero-point according to \citet{2021A&A...649A...4L}. We find that these additional features do not only help to estimate \logg, but they also improve our \MH\ estimates by $\sim$10\% where metal-poor giants benefit in particular. The complete list of all input features and the details of the XGBoost configuration are provided in Appendix~\ref{appendix:details-XGBoost}.

We use the exact same set of input features for training the XGBoost models for \MH, \Teff\ and \logg, all based on training labels (see Sect.~\ref{ssec:training-sample}). Our objective is to maximize the number of stars for which all required input features are available. In that case, the completeness of our results would be dominated by the completeness of AllWISE photometry (see Fig.~\ref{fig:completeness_vs_GBP}).

\subsection{Internal 20-fold cross validation}

\begin{figure*}
\begin{center}
\includegraphics[width=2\columnwidth]{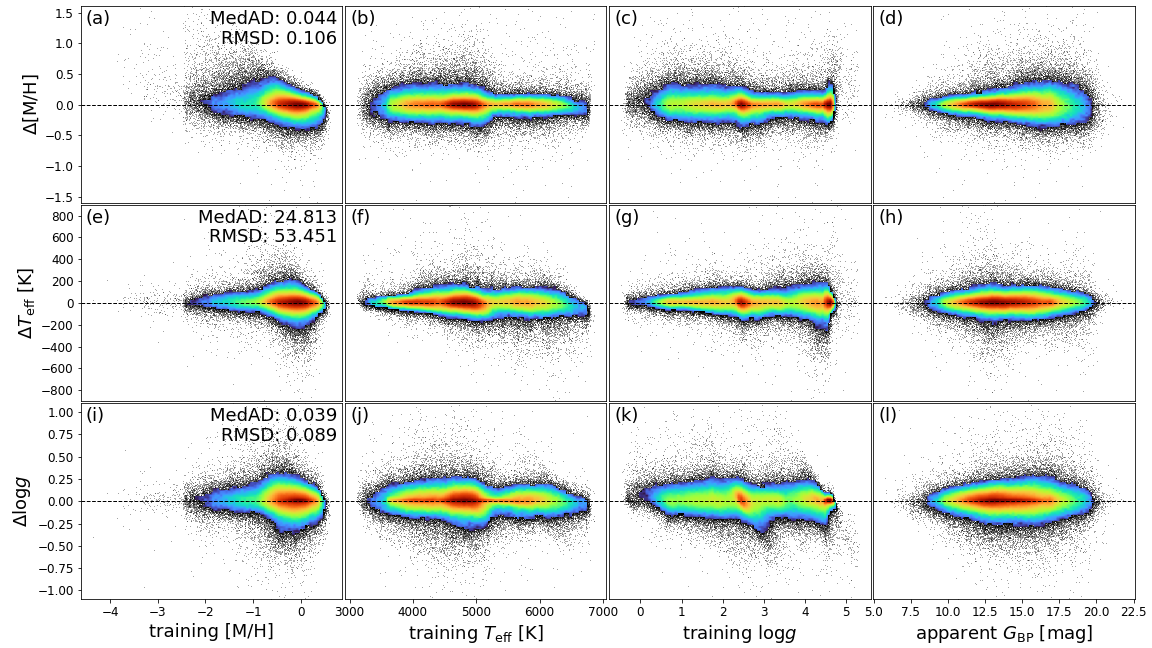}
\end{center}
\caption{Cross-validation of the XGBoost parameters on the training sample (SDSS-APOGEE~DR17 and \citet{2022ApJ...931..147L}). The plots show the results of the twenty-fold cross-validation of the 5\% portions of the training sample, withheld in the training. Rows from top to bottom show \MH\ residuals, \Teff\ residuals and \logg\ residuals. Columns from left to right show residuals vs.\ the training sample's \MH, \Teff, \logg\ and Gaia's apparent $G_\textrm{BP}$ magnitude. The numbers in the top right corners quote the median absolute difference (MedAD) and the root mean square difference (RMSD). The density map is logarithmic. These plots illustrate the remarkable precision of the approach: 0.10~dex in \MH, 50K in \Teff , and 0.08~dex in \logg . Note that these variances still include all the uncertainties in the APOGEE estimates.}
\label{fig:internal-x-validation-XGBoost}
\end{figure*}

For internal validation, we assess the quality of XGBoost results on the training sample, using 20-fold cross-validation: 20 times we set aside disjoint sets comprising 5\% of the data for subsequent testing of a model trained on the other 95\% of the data. In the end, the data features for each object in the training sample have been compared to a statistically independent XGBoost model prediction for them.  These cross-validation results are summarized in Fig.~\ref{fig:internal-x-validation-XGBoost}. 

\begin{figure}
\begin{center}
\includegraphics[width=\columnwidth]{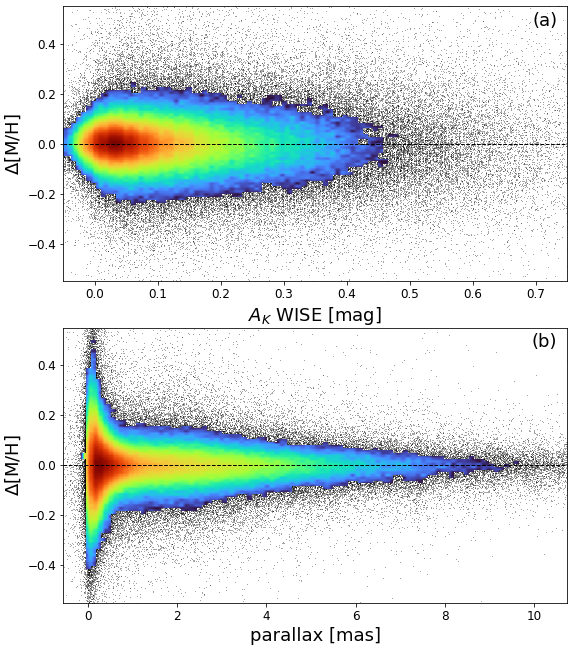}
\end{center}
\caption{Cross-validation of XGBoost on the training sample (SDSS-APOGEE~DR17 and \citet{2022ApJ...931..147L}). Dependence of test error for \MH\ on WISE $A_K$ extinction (panel a) and parallax (panel b). \MH\ shows no systematics with either.}
\label{fig:internal-x-validation-XGBoost-vs-AK-and-parallax}
\end{figure}

The first row of Fig.~\ref{fig:internal-x-validation-XGBoost} is most important because it shows the cross-validation of the \MH\ estimates. Panel (a) shows that, for the most part, our results are accurate, i.e.\ unbiased with respect to the APOGEE reference \MH. There are only a few outliers where XGBoost assigns a higher \MH\ value than APOGEE. However, for training labels $\MH<-3$ XGBoost tends to overestimate \MH. Most likely, this is a consequence of mixing different definitions of ``metallicity'' in our training sample: While APOGEE provides \MH\ estimates, \citet{2022ApJ...931..147L} provide estimates of [Fe/H]. Astrophysically, very old stars will have low Iron content but may already have been enhanced in other elements, such that the stars from \citet{2022ApJ...931..147L} may be genuinely low in [Fe/H] but have higher \MH, which is recognized by XGBoost learning \MH\ from the majority of training examples provided by APOGEE. Panel (b) shows how our \MH\ estimates depend on \Teff\ values: the agreement is overall very good, and for stars hotter than $\sim$5500K our \MH\ estimates are closer to the training labels than for cooler stars. The main reason for this is that for $\Teff>5500$K the APOGEE sample contains virtually no stars with \MH\ below -0.75, limiting the comparison to the ``easy'' metral-rich regime. Panel (c) shows that our \MH\ residuals do not exhibit any noteworthy trends with the training sample's \logg, i.e.\ our \MH\ estimates work just as good for main-sequence dwarfs as they work for red giant stars. Panel (d) shows that our \MH\ estimates remain robust (though obviously less precise) to the very faint end of $G_\textrm{BP}\sim 20$. The overall root-mean-square difference to the training sample's \MH\ estimates is 0.106 and half of the stars differ by less than 0.044 from their reference values. This performance is about 10\% better than that found for the sample of only red giants in \citet{Rix2022} (see Table~1 therein). This improvement, despite the expanded coverage of the CMD, is mainly due to the inclusion of luminosity estimates (see Eq.~(\ref{eq:feature-with-parallax})) as features.
Like in \citet{Rix2022}, the current \MH\ estimates remain unbiased as the $A_K$ extinction increases as is evident from Fig.~\ref{fig:internal-x-validation-XGBoost-vs-AK-and-parallax}a. Including CatWISE photometry is the key here. Furthermore, 
Fig.~\ref{fig:internal-x-validation-XGBoost-vs-AK-and-parallax}b establishes that there are also no systematics with parallax (i.e.\ inverse distance).

In particular, \citet{Rix2022} restricted their analysis to bright ($G_\textrm{BP}<16$mag) red-giant-branch (teff\_xgboost$<$5500K and logg\_xgboost$<$3.5) stars, Fig.~\ref{fig:internal-x-validation-XGBoost} panels (b) and (c) suggest that the \MH\ estimates from the current work are also robust outside the RGB and panel (d) suggests that this also holds down to the faintest stars which have their XP spectra published in \gdrthree. Note that we cannot test with this validation sample whether our \MH\ estimates remain so precise and robust also for very metal poor stars.

The other two rows of Fig.~\ref{fig:internal-x-validation-XGBoost} show the XGBoost residuals for \Teff\ (middle) and \logg\ (bottom). The RMS differences are remarkably small: 54K for \Teff\ and 0.089 for \logg . Furthermore, the residuals do not show obvious systematics and appear to remain robust down to $G_\textrm{BP}\sim 20$.

\section{Stellar Parameters from XP Spectra via XGBoost}
\label{sec:application}

We now turn to applying the XGBoost estimator, trained as just described, to an all-sky sample of stellar sources with XP spectra and CatWISE photometry.

\subsection{Sample selection}

\begin{figure}
\begin{center}
\includegraphics[width=\columnwidth]{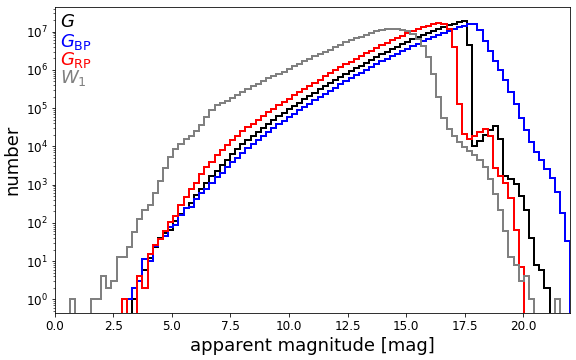}
\end{center}
\caption{Apparent magnitude distributions --- $G$ (black), $G_\textrm{BP}$ (blue), $G_\textrm{RP}$ (red) and $W_1$ (grey) -- for the full application sample of 174\,922\,161 stars without any quality cuts. The bump in the $G$-band distribution at $G\sim 19$ reflects contamination by galaxies, QSOs, and ultracool dwarfs.}
\label{fig:distribution-apparent-magnitudes}
\end{figure}

We define the sample to which we apply the XGBoost estimator as all sources in Gaia DR3 that have XP spectra, valid parallaxes and proper motions, and valid XP-derived and CatWISE photometry; the parallaxes do not have to differ significantly from zero. The following AQDL query
\begin{verbatim}
SELECT 
source_id
FROM gaiadr3.gaia_source_lite
WHERE has_xp_continuous='true' 
AND parallax IS NOT NULL
\end{verbatim}
results in 218\,132\,063 stars. Since we require complete photometry in CatWISE and synthesized passbands (see Sect.~\ref{ssec:completeness-photometry}), not all of them have the complete set of input features to XGBoost. The final number of stars that satisfy these additional conditions is 174\,922\,161 ($\sim$80.2\%). Their apparent magnitude distributions are shown in Fig.~\ref{fig:distribution-apparent-magnitudes}. It is important to note that in Gaia DR3 XP spectra were only published for sources brighter than $G=17.65$, yet Fig.~\ref{fig:distribution-apparent-magnitudes} shows sources fainter than that. The reason is that the XP spectra of presumed QSOs, galaxies, and ultracool dwarfs were exempt from the Gaia DR3 publication limit of $G=17.65$. Consequently, these objects may be contaminants in our stellar parameter catalog: they manifest as a small bump at $G\sim 19$ in the distributions of $G$ and $G_\textrm{RP}$. 

The overall result of this analysis is given in Table~\ref{table:full-results}: the three stellar parameters, \MH, \Teff, \logg\ for 175 million sources, specified by their \gdrthree\ source~ID and with a label whether these sources were included in the XGBoost training.

\subsection{External validation with other surveys}

We can validate these XGBoost results by a comparison to other surveys not used in the training. Specifically, we compare to results from GSP-Spec after calibration\footnote{\url{https://www.cosmos.esa.int/web/gaia/dr3-gspspec-metallicity-logg-calibration}} in Gaia DR3 \citep{Recio-Blanco-GSPspec}, GALAH DR3 \citep{GALAH_DR3}, and SkyMapper DR2 \citep{2021ApJS..254...31C}.

\begin{figure*}
\begin{center}
\includegraphics[width=2\columnwidth]{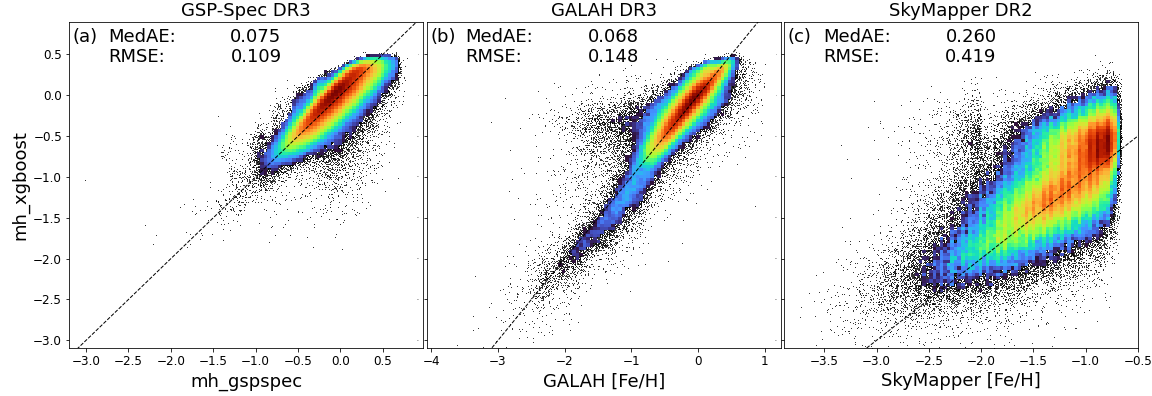}
\end{center}
\caption{Comparison of XGBoost \MH\ estimates with GSP-Spec's calibrated metallicity estimates in Gaia DR3 \citep{Recio-Blanco-GSPspec}, with GALAH DR3 \citep{GALAH_DR3}, and with SkyMapper DR2 \citep{2021ApJS..254...31C}. For Skymapper DR2, we impose the quality flag equal to 0. Numbers quote the median absolute difference (MedAD) and the root-mean-square difference (RMSD). No quality cuts were applied, and the density maps are logarithmic. The comparison with GSP-Spec DR3 is very good, but the comparison is limited (mostly by GSP-Spec) to $\MH\gtrsim -1$. The comparison with GALAH DR3 is also very good, with a small systematic offset at $\MH\le -1$, already noted in \citet{Rix2022}. Comparison with the photometric \MH\ estimates from SkyMapper~DR2 shows a substantially increased scatter. The good comparison of XGBoost results with other surveys makes it likely that this is attributable to SkyMapper issues.}
\label{fig:external-validation-mh_xgboost}
\end{figure*}

To start, we compare \MH\ estimates from XGBoost with other metallicity estimates. For GSP-Spec (Fig.~\ref{fig:external-validation-mh_xgboost}a), the agreement is excellent: there are no discernable systematics and very few outliers. Importantly, the GSP-Spec comparison is mostly limited to $\MH > -1$, where we expect \MH\ estimates to be robust. For GALAH (Fig.~\ref{fig:external-validation-mh_xgboost}b), we still see a good overall agreement across the full metallicity range. However, there are some outliers, where GALAH estimates [Fe/H] below $-1$, while XGBoost estimates \MH\ above $-0.5$. We also note a small systematic offset below [Fe/H] of $-1$ where XGBoost's \MH\ is $\sim$0.2 lower than GALAH's [Fe/H]. These outliers and the slight offset have also been observed in the results of \citet{Rix2022}. Yet, unlike in \citet{Rix2022}, we no longer see a saturation of XGBoost metallicities below $-2$, where we now see a continuation of the one-to-one relation with GALAH. This is the result of including the very metal-poor stars of \citet{2022ApJ...931..147L} in our training sample, thus extending the APOGEE metallicity range. For the SkyMapper photometric metallicities (Fig.~\ref{fig:external-validation-mh_xgboost}c), a substantial scatter is evident both visually and quantitatively. Successful comparisons to external spectroscopic surveys suggest that this scatter is probably inherent to SkyMapper.

\begin{figure}
\begin{center}
\includegraphics[width=\columnwidth]{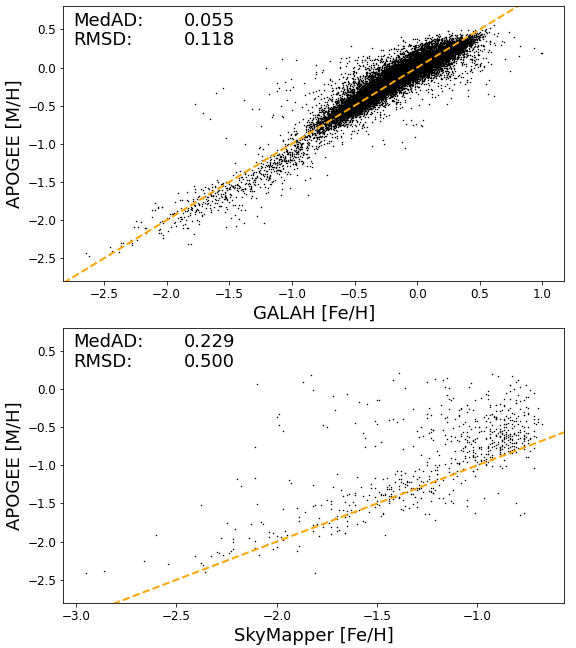}
\end{center}
\caption{Comparison of \MH\ estimates from APOGEE DR17 \citep{SDSS_DR17} to [Fe/H] estimates from GALAH DR3 \citep{GALAH_DR3} (top panel) and to [Fe/H] estimates from SkyMapper DR2 \citep{2021ApJS..254...31C} (bottom panel). There are 22\,662 stars in common between APOGEE and GALAH and 696 in common between APOGEE and SkyMapper. Numbers quote the median absolute difference (MedAD) and the root-mean-square difference (RMSD).}
\label{fig:GALAH-vs-APOGEE-metals}
\end{figure}

We further investigate the origin of the outliers and systematics of our \MH\ estimates in Fig.~\ref{fig:GALAH-vs-APOGEE-metals} where we directly compare metallicity estimates from APOGEE (i.e.\ the training sample underlying XGBoost) to those from GALAH: First, we observe the same small systematic offset below [Fe/H] of $-1$ between GALAH and APOGEE, i.e.\ the XGBoost model has correctly learned from APOGEE and simply reflects this difference. Second, we can also see the outliers where GALAH vote for ${\rm [Fe/H]}<-1$ whereas APOGEE votes for $\MH>-0.5$, so again XGBoost has faithfully learned from its APOGEE training sample. Consequently, both effects are traced back to genuine differences between GALAH and APOGEE and are thus not introduced by XGBoost. In fact, these outliers in Fig.~\ref{fig:external-validation-mh_xgboost}b mostly have high temperatures in GALAH and potentially correspond to outliers in GALAH DR3 itself.

Quantitatively, Fig.~\ref{fig:external-validation-mh_xgboost} shows that our XGBoost \MH\  estimates compare very well with those from GSP-Spec and GALAH, with half of the stars differing by no more than 0.092 and 0.068, respectively. This external validation error is somewhat larger than the cross-validation error of 0.042 found for APOGEE in Fig.~\ref{fig:internal-x-validation-XGBoost}. This most likely reflects subtle differences between APOGEE's and GSPSpec's \MH\ estimates (our XGBoost estimates are tied to the APOGEE scale), as a similar scatter is found in the direct comparison of these surveys.

For the scatter of temperatures and surface gravities when comparing with GSP-Spec we find RMS differences of 112K for \Teff\ and 0.223 for \logg; and when comparing with GALAH, we find 166K for \Teff\ and 0.119 for \logg, respectively. These are again slightly higher than the RMS differences from the 20-fold cross-validation on APOGEE (54K and 0.089, respectively) quoted in Fig.~\ref{fig:internal-x-validation-XGBoost}. For \logg, the difference to GSP-Spec is larger than for APOGEE or GALAH, but we also did not apply the empirical corrections for GSP-Spec's \logg\ recommended in \citet{Recio-Blanco-GSPspec}.

\subsection{\MH\ estimates at faint apparent magnitudes}

\begin{figure}
\begin{center}
\includegraphics[width=\columnwidth]{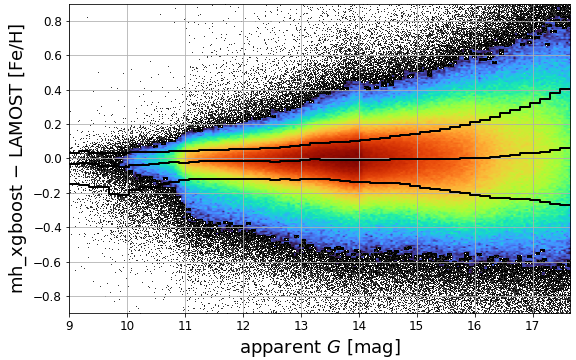}
\end{center}
\caption{Differences of \MH\ between XGBoost for LAMOST DR6 as function of apparent $G$ magnitude. The Gaia DR3 publication limit for XP spectra is $G=17.65$. Black lines indicate the 16th, 50th and 84th percentiles as function of $G$. Color maps indicate logarithmic number density.}
\label{fig:MH-residuals-vs-apparent-G}
\end{figure}

Of particular interest is the publication limit of XP spectra in Gaia DR3, which was set at $G=17.65$. Figure~\ref{fig:internal-x-validation-XGBoost}d suggests that our XGBoost results may remain robust as we approach the publication limit, but we would like to confirm this with an independent validation sample. Unfortunately, both GSP-Spec and GALAH DR3 are of no use in exploring this regime, as both samples are limited to bright stars. Therefore, we make use of the LAMOST DR6\footnote{\url{https://dr6.lamost.org/}} data \citep{2011RAA....11..924W,2014IAUS..306..340W}. As is evident from Fig.~\ref{fig:MH-residuals-vs-apparent-G}, the \MH\ differences between XGBoost and LAMOST degrade ``gracefully'' towards the faint end, which means that the random scatter increases smoothly and no systematics appear. At $G=16$ the central 68\% interval ranges from $-0.2$ to $+0.2$ and even at $G=17.65$ it ranges from $-0.3$ to $+0.4$. In fact, these variances include the [Fe/H] uncertainties from LAMOST, which typically are of the order of 0.25 around $G=17.65$. Assuming that these uncertainties add in quadrature, the $-0.3$ to $+0.4$ interval at $G=17.65$ implies an uncertainty of 0.17 -- 0.32 attributable to XGBoost.

While a random error of $\sim 0.33$ in \MH\ at $G\sim 17$ is acceptable, it still represents a substantial increase from the error of 0.1 at the bright end. What are the possible origins of this increased noise? First, an earlier version of our catalog was based on AllWISE photometry instead of CatWISE, but apart from having significantly lower completeness (see Fig.~\ref{fig:completeness_vs_GBP}) it produced the same results at the faint end. This rules out the CatWISE photometry as origin for the increased noise. Second, \gdrthree\ parallaxes can become very noisy towards the faint end, such that the features defined in Eq.~(\ref{eq:feature-with-parallax}) could begin to confuse XGBoost at the faint end. However, if we remove these features from the XGBoost input and thus become entirely independent from the parallax, we find no improvement either. This only leaves the XP spectra as the source of the increased noise towards the faint end. More precisely, we suspect that it is not the XP coefficients themselves but rather the synthesized narrow-band photometry which is becoming increasingly susceptible to noise towards the faint end. This interpretation is also supported by Fig.~\ref{fig:completeness_vs_GBP}, which reminds us that the synthetic photometry becomes incomplete due to noise leading to negative flux values even when XP spectra are available.


\subsection{External validation with solar analogs}

\begin{figure}
\begin{center}
\includegraphics[width=\columnwidth]{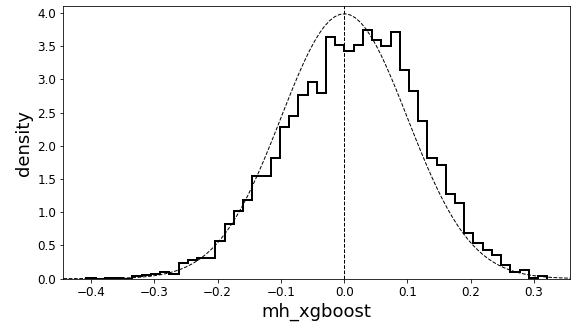}
\end{center}
\caption{Distribution of \MH\ estimates from XGBoost for 5759 Solar analog candidates from \citet{2022arXiv220605870G}. A Gaussian with zero mean and standard deviation of 0.1 is given by the dashed line, illustrating that the \MH\ estimates are precise and accurate on the main sequence at high metallicities and for $\Teff\sim 5772$K.}
\label{fig:MH-distribution-solar-analogues}
\end{figure}

\citet{2022arXiv220605870G} compiled a list of 5863 Solar-analog candidates, whereof 5759 are in our sample. According to XGBoost, their mean \MH\ is $0.012\pm 0.105$ and the central 90\% interval ranges from -0.167 to 0.178. Figure~\ref{fig:MH-distribution-solar-analogues} shows their distribution, which is consistent with a Gaussian of standard deviation 0.1. This is in excellent agreement with the Solar value and demonstrates that our \MH\ estimates are also reliable at least for Solar-like main-sequence dwarfs, whereas the estimates from \citet{Rix2022} were applicable only to giant stars.

\subsection{External validation with clusters}
\label{ssec:validation-with-clusters}

\begin{figure}
\begin{center}
\includegraphics[width=\columnwidth]{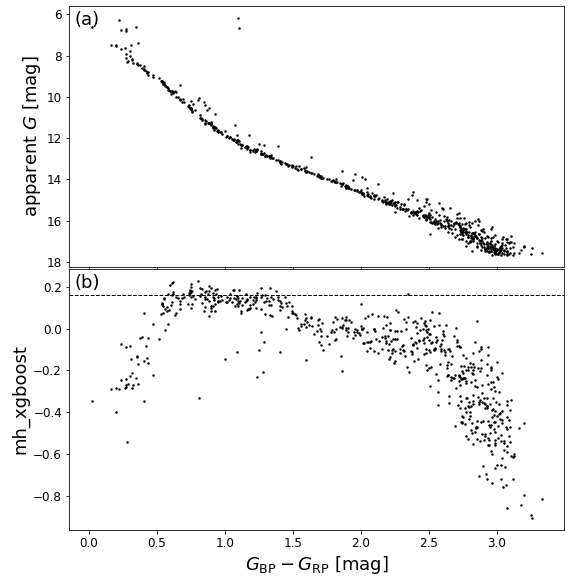}
\end{center}
\caption{Validation of the \MH\ estimates in the main sequence, using the Praesepe cluster. The cluster's color-magnitude diagram of all 653 members with \MH\ is illustrated in panel (a). Their \MH\ estimates are shown as a function of color in panel (b). The horizontal dashed line indicates the metallicity of 0.16 ($Z=0.02$) adopted by \citet{2018A&A...616A..10G}. For $0.5<G_\textrm{BP}-G_\textrm{RP}<1.5$ the metallicity agreement is excellent, whereas for $1.5<G_\textrm{BP}-G_\textrm{RP}<2.5$ they are systematically too low by 0.15~dex. Outside of these color-ranges the agreement is poor. We attribute the offsets and the poor estimates to possible systematics and poor sampling of the CMD space in the training sample. The \MH\ estimates for main-sequence stars with colors outside $0.5<G_\textrm{BP}-G_\textrm{RP}<2.5$ are manifestly unreliable.}
\label{fig:Praesepe-MH-vs-colour}
\end{figure}

Among our XGBoost results, we find 22\,477 member stars in 36 open clusters from \citet{2018A&A...616A..10G}. As a first instructive example, Fig.~\ref{fig:Praesepe-MH-vs-colour}b shows how XGBoost's metallicity estimate varies with $G_\textrm{BP}-G_\textrm{RP}$ color in the Praesepe cluster. The expected literature value is recovered only within a certain color range but otherwise XGBoost systematically underestimates the metallicity. This underestimate is related to the limited temperature range of the training sample (3107K -- 6867K, see Sect.~\ref{ssec:training-sample}). In the absence of interstellar extinction (such as for Praesepe), this temperature range roughly corresponds to a $G_\textrm{BP}-G_\textrm{RP}$ color range from 0.5 to 2.3. Indeed, Fig.~\ref{fig:Praesepe-MH-vs-colour}b shows that the underestimated \MH\ occurs mainly for colors bluer than 0.5 or redder than 2.5, with a less pronounced underestimation by about 0.15 also in the range from 1.5 to 2.5. As is also evident from Fig.~\ref{fig:Praesepe-MH-vs-colour}a, the Praesepe member stars are dominated by main-sequence dwarfs. We also note that the Solar analogs showing excellent agreement in Fig.~\ref{fig:MH-distribution-solar-analogues} have intrinsic colors of $G_\textrm{BP}-G_\textrm{RP}=0.818 \pm 0.029$ \citep{2022arXiv220605870G} and would thus fall well within the regime of good agreement between 0.5 and 1.5 in Fig.~\ref{fig:Praesepe-MH-vs-colour}b.

\begin{figure}[h]
\includegraphics[width=\columnwidth]{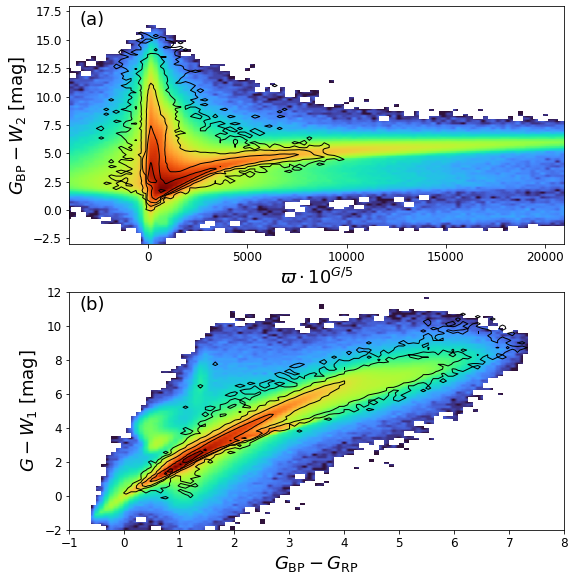}
\caption{XGBoost input feature distributions of the application sample (colormaps), overlayed with black contours of the training sample. The lowest contour is at 1 star per bin, i.e.\ it encloses the full training sample, and the other contours successively increase by factors of 10. This shows that our full sample extends across important portions of color-color space that are not covered by the training sample. The resulting parameter estimates in these regimes will be inevitably unreliable.}
\label{fig:colour-colour-distribution-application-vs-training-sample}
\end{figure}

\begin{figure}
\begin{center}
\includegraphics[width=\columnwidth]{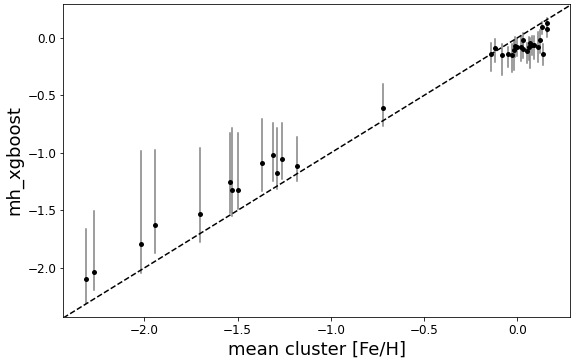}
\end{center}
\caption{Comparison of \MH\ estimates from XGBoost after filtering on input features to mean metallicities of 36 open clusters from \citet{2018A&A...616A..10G}. Black dots show median \MH\ and grey errorbars show 16th and 84th percentiles in each cluster.}
\label{fig:validation-open-clusters}
\end{figure}

Obviously, many Praesepe member stars fall into a temperature range that is not covered sufficiently by our training sample. Unfortunately, we cannot select on our catalog's XGBoost temperature estimate because its values are also strictly confined to the range 3107K -- 6867K of training examples. Instead, we select on input features as shown in Fig.~\ref{fig:colour-colour-distribution-application-vs-training-sample}: for every individual cluster member star, we ask where its features fall into the two diagrams and we reject it from further consideration if and only if it has a sufficiently high number of training examples nearby in these diagrams.\footnote{The exact thresholds for this selection depend on the binning used in Fig.~\ref{fig:colour-colour-distribution-application-vs-training-sample}.}
After this filtering procedure, Fig.~\ref{fig:validation-open-clusters} shows that the XGBoost \MH\ estimates agree reasonably well with the adopted mean metallicities of the 36 open clusters from \citet{2018A&A...616A..10G}. We do see a slight positive offset below [Fe/H] of -1 and a slight negative offset around solar [Fe/H]. The latter is probably similar to the offset seen in Fig.~\ref{fig:Praesepe-MH-vs-colour} for Praesepe and colors $G_{\rm BP}-G_{\rm RP}>1.4$. The slight positive offset below [Fe/H] of -1 may be due to XGBoost occasionally overestimating \MH\ in that regime (see Fig.~\ref{fig:internal-x-validation-XGBoost}a).

\subsection{External validation with wide binaries}
\label{ssec:wide-binaries}

\begin{figure}
\begin{center}
\includegraphics[width=\columnwidth]{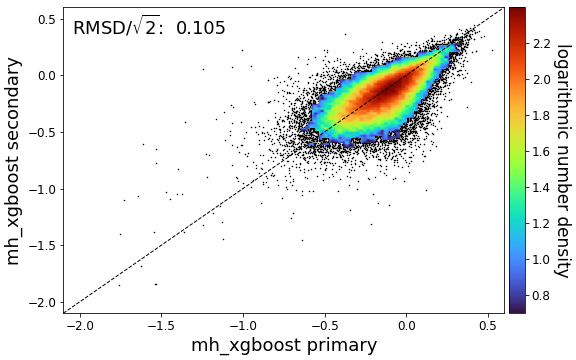}
\end{center}
\caption{Comparison of \MH\ estimates for components of 30\,748 wide binaries from \citet{WideBinaries}. We quote the root-mean-squared difference (RMSD) divided by $\sqrt{2}$ because we want to quantify the difference between two noisy \MH\ estimates.}
\label{fig:external-validation-wide-binaries}
\end{figure}

Given the catalog of \citet{WideBinaries}, we find 55\,033 pairs of wide binaries where each component star is observed as an individual source by Gaia. Since both stars from each binary pair have formed from the same gas cloud, XGBoost should estimate the same metallicity. In fact, Fig.~\ref{fig:external-validation-wide-binaries} shows that their \MH\ estimates are consistent with each other. The RMS difference divided by $\sqrt{2}$ is 0.105.

An inspection of the XGBoost surface gravities reveals that essentially all wide binary members are main-sequence dwarf stars, which most likely is a selection effect from \citet{WideBinaries} focusing on high-quality astrometry in Gaia DR2. Being main-sequence stars, the limitations from Fig.~\ref{fig:Praesepe-MH-vs-colour}b apply: If we restrict the comparison to wide binaries in which both stars fall within the color range $0.5<G_\textrm{BP}-G_\textrm{RP}<1.5$, the RMS difference divided by $\sqrt{2}$ drops from 0.105 to 0.079 and the mean difference is 0.004. In contrast, if we only consider pairs where one component is in the good color range, whereas the second component is within $1.5<G_\textrm{BP}-G_\textrm{RP}<2.5$, the redder stars have on average 0.05 lower \MH\ than the (unbiased) bluer stars. This offset is slightly less than the systematic underestimation of -0.15 seen in Praesepe in Fig.~\ref{fig:Praesepe-MH-vs-colour}b.

We also note that while most of the wide binaries in Fig.~\ref{fig:external-validation-wide-binaries} are at $\MH>-0.5$, there are a handful of systems with $\MH<-1$ and that in those cases the XGBoost estimates still hold.

\subsection{OBA stars as a failure mode}
\label{ssec:failure-modes}

In this section, we investigate the results for OBA stars. \citet{2022arXiv220605870G} compiled a list of 3\,023\,388 OBA stars, whereof 2\,371\,118 are in our sample. These are beyond the temperature range of the training sample, i.e.\ we intentionally break the model assumptions of our XGBoost model. Given that absorption lines are often washed out in very hot stars, we expect XGBoost to misinterpret such stars as metal-poor. Since OBA stars are mostly young, they should have Solar-like metallicities or higher. However, XGBoost assigns a median \MH\ of $-0.443$ to the stars in this sample and about 11\% of OBA stars are assigned a \MH\ lower than $-1$ by XGBoost. Consequently, the XGBoost estimates are clearly not viable for OBA stars. We note, however, that \citet{2022arXiv220605870G} report contamination from metal-poor stars, i.e.\ not all of those may actually be hot OBA stars.

\begin{table*}
\centering
\caption{Abridged table of 174\,922\,161 XGBoost estimates presented in this work. The full table is available online \citep{Andrae2023Zenodo}. The Gaia DR3 \texttt{source\_id} is sorted in ascending order. The boolean flag \texttt{in\_training\_sample} indicates whether or not a source was part of the XGBoost training sample. We provide minimal information in order to save data volume.}
\label{table:full-results}
\begin{tabular}{c|c|c|c|c|c|c}
source\_id & catwise\_w1 & catwise\_w2 & in\_training\_sample & mh\_xgboost & teff\_xgboost & logg\_xgboost \\
\hline
4295806720 & 15.796 & 15.942 & False & -0.256 & 5991.6 & 4.551 \\
38655544960 & 11.837 & 11.879 & False & -0.212 & 4791.6 & 4.604 \\
1275606125952 & 14.366 & 14.438 & False & -0.438 & 5177.3 & 4.489 \\
1653563247744 & 14.734 & 14.802 & False & -1.286 & 6102.0 & 4.017 \\
2851858288640 & 10.904 & 10.932 & True & -0.454 & 5899.5 & 4.295 \\
3332894779520 & 10.316 & 10.385 & True & 0.178 & 4915.7 & 3.594 \\
3371550165888 & 12.280 & 12.350 & False & -0.388 & 4912.7 & 4.520 \\
3508989119232 & 13.386 & 13.439 & False & -0.518 & 5339.2 & 4.558 \\
4711579935744 & 12.725 & 12.767 & False & -0.323 & 5896.7 & 4.338 \\
4814659150336 & 14.207 & 14.241 & False & -0.289 & 4163.3 & 4.667 \\
\vdots & \vdots & \vdots & \vdots & \vdots & \vdots & \vdots
\end{tabular}
\end{table*}

\section{Illustration of the Sample}
\label{sec:illustration}

The results of our XGBoost analysis are listed in Table~\ref{table:full-results}. They are unprecedented in sample size at such precision and accuracy ($\sigma (\MH ) \sim 0.1$~dex, $\sigma (\Teff) \sim 50$K, $\sigma (\logg ) \sim 0.08$~dex) and can be used for a vast array of science applications, which is beyond the scope of this paper.  This combination of sample size and data quality warrants a rigorous modeling of the selection function \citep[see e.g.][]{Rix2021}, which are also beyond the scope of this paper. 
What we will do here is to provide two, only qualitative, illustrations of the sample's science potential: its total \MH\ distribution, and all-sky maps in different bins of \MH. More generally, we emphasize that each science application warrants specific vetting of the subsample used.

\subsection{\MH\ distribution of the sample}

\begin{figure}[h]
\includegraphics[width=\columnwidth]{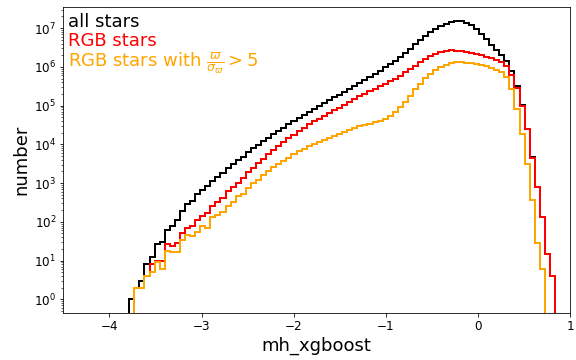}
\caption{\MH\ distributions resulting from XGBoost for all stars in our catalog (black), RGB stars (red, teff\_xgboost$<$5300K and logg\_xgboost$<$3.5) and RGB stars with high-quality parallaxes (orange). The differences in the distributions between the RGB's with and without high-quality parallaxes is physical, as the parallax cut eliminates mostly distance stars, often in the halo or the Magellanic Clouds, which are metal-poor. The parallax cut on the RGB sample matters, as most of this sub-sample also has RVS radial velocities: for these stars orbits can be calculated \citep[e.g.][]{Rix2022}.}
\label{fig:sample-MH-distribution}
\end{figure}

Perhaps the most compact way to present the sample is to show its \MH\ distribution. We show this distribution in Fig.~\ref{fig:sample-MH-distribution} for all 174\,922\,161 stars, for 43\,520\,755 likely RGB stars, and for 18\,858\,968 RGB stars with high-quality parallaxes (see below). The \MH\ distribution for this last subset in  Fig.~\ref{fig:sample-MH-distribution}, restricted to the Milky Way within about 10~kpc, is the one that can be taken most at face value as an observational approximation of the ``total'' metallicity distribution of the Galaxy. Of course this is a flux-limited sample, whose extent is limited by distance, dust extinction and (in part of the sky) crowding. Proper volume corrections of this \MH\ distribution would be a complex exercise \citep[see, e.g.][]{Rix2021} beyond the scope of this work.  Such an analysis must include the $G\leq 17.65$ publication limit of XP spectra in Gaia DR3 (e.g.\ introducing foreground dust extinction), the completeness of photometry synthesized from XP spectra (i.e.\ loss of stars due to negative synthetic fluxes caused by noisy XP spectra, see Sect.~\ref{ssec:completeness-photometry}) and the crowding-afflicted completeness of AllWISE photometry.

Nonetheless, this distribution shows a number of remarkable features, extending over a factor of 10\,000 in metallicity within a single galaxy. It starts at $\MH\approx -3.5$, rises steeply to $\MH\approx -2.5$, and then follows $d\log N/d\MH \sim 1$ to  $\MH\approx -1.0$. At this point, the onset of the old disk in metallicity, the \MH\ distribution rises quickly to a maximum near $\MH\approx -0.4$, stays flat to $\MH\approx +0.2$, dropping steeply beyond. The implications of this distribution in terms of chemical enrichment warrant to be studied in a framework of chemical evolution models, such as that of \citet{Weinberg2017}. The interpretation in terms of halo, old disk, thin disk, etc., will be most powerful when combining this information with orbital information. We do not pursue these avenues in this paper, but stress only two points: first, our training set, extended in \MH\ compared to \citet{Rix2022}, shows that there are likely hundreds of mostly bright ($G<16$) extremely metal-poor giants ($\MH\ge -3$) and over 10\,000 very metal-poor giants ($\MH\ge -2$) observable in the Milky Way. Second, the steep slope of the \MH\ distribution argues that spurious \MH\ determinations remain rare also at the lowest accessible metallicities.

\subsection{Mono-abundance, all-sky maps}

\begin{figure}[h]
\includegraphics[width=\columnwidth]{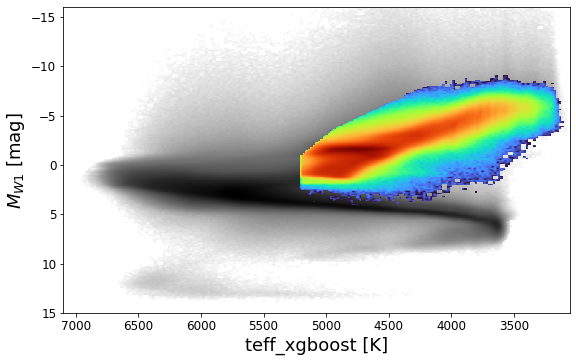}
\caption{Illustration of quality cuts in effective temperature and $M_{W1}=W_1+5\cdot\log_{10}(\varpi/100)$ for the vetted RGB sample (colored density map, 17\,558\,141 stars) compared to the full sample (gray density map, 174\,922\,161 stars). The cuts were designed with two goals in mind: first, isolate a subsample of bright giants for which the \MH\ estimates should be most precise and robust, to be used e.g.\ in Galactic chemodynamics. Second, they are limited in temperature to $5200$K, which was empirically found to be highly effective to eliminate contamination of the metal-poor subsample by unrecognized hotter and reddened stars.}
\label{fig:clean-RGB-cuts}
\end{figure}

\begin{figure*}
\centering
\begin{minipage}[t]{\columnwidth}
\begin{center}
\includegraphics[width=\columnwidth]{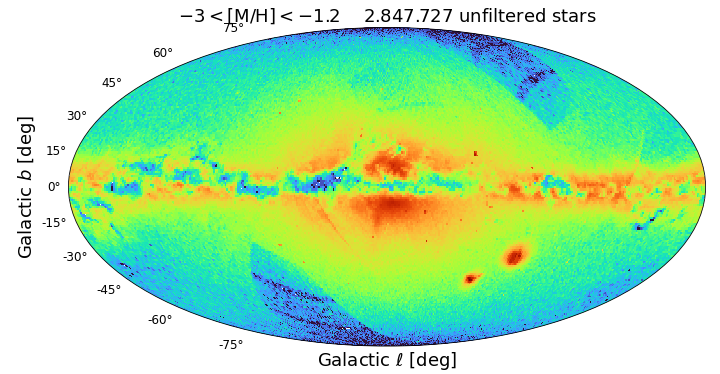}
\includegraphics[width=\columnwidth]{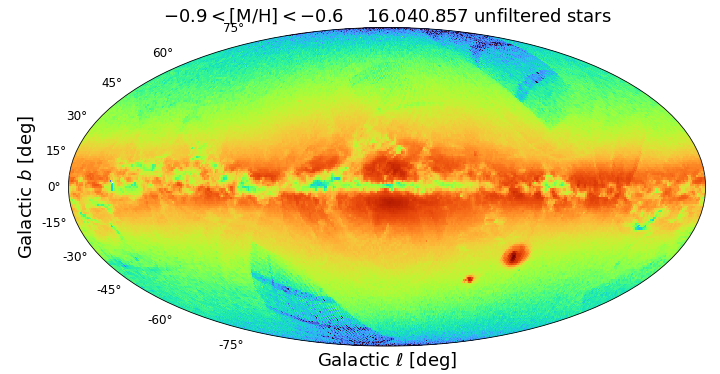}
\includegraphics[width=\columnwidth]{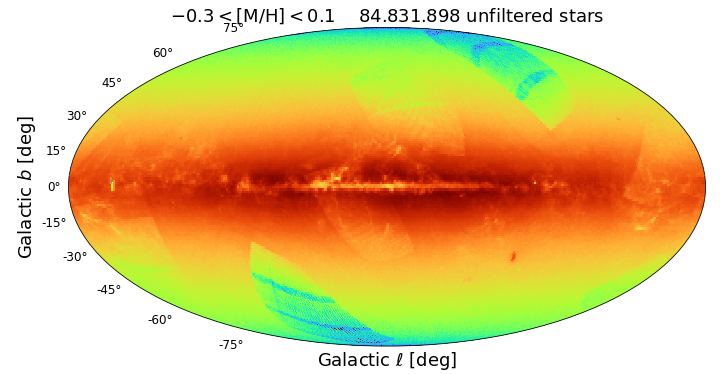}
\end{center}
\caption{Skymaps showing the logarithmic number density of all \emph{unfiltered} stars with $-3<\MH<-1.2$ (top panel), $-0.9<\MH<-0.6$ (middle panel) and $-0.3<\MH<0.1$ (bottom panel). These illustrate two important issues: the incompleteness of XP spectra in the two sickle-shaped regions at high latitude. And a significant contamination of the metal-poor bin in the unfiltered sample, which manifests itself as a thin disk near the Galactic plane; these are presumably hotter and highly reddened stars with weaker metal lines that are not recognized as such by our algorithm, as it lacks good training sets in this CMD regime.}
\label{fig:mono-abundance-all-sky}
\end{minipage}
\hspace{0.5cm}
\begin{minipage}[t]{\columnwidth}
\begin{center}
\includegraphics[width=\columnwidth]{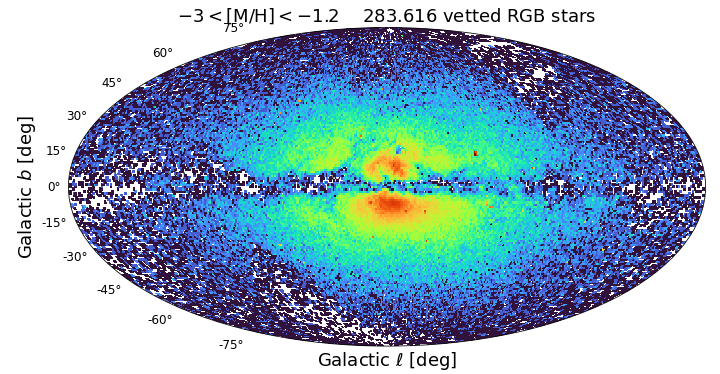}
\includegraphics[width=\columnwidth]{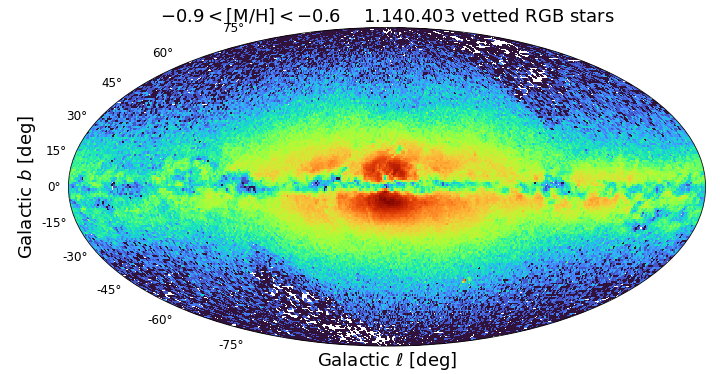}
\includegraphics[width=\columnwidth]{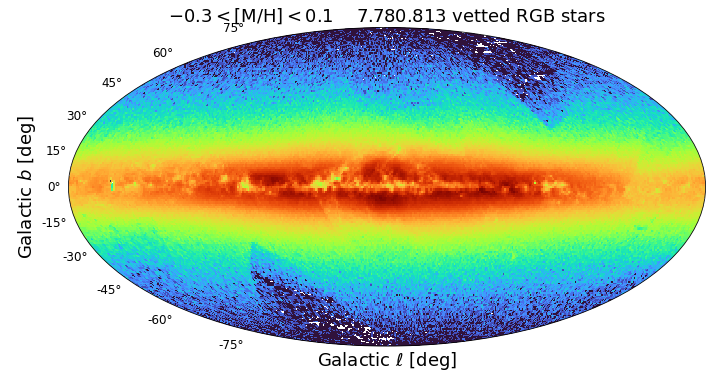}
\end{center}
\caption{Skymaps showing the logarithmic number density of \emph{vetted} RGB stars for $-3<\MH<-1.2$ (top panel), $-0.9<\MH<-0.6$ (middle panel) and $-0.3<\MH<0.1$ (bottom panel). This vetted sample (see Fig.~\ref{fig:clean-RGB-cuts}) is restricted to red giants with good S/N ($G<16$), \Teff\ cuts that eliminate contaminants in the low-metallicity subsample and significant parallax (which explains the ``disapperance'' of the Magellanic Clouds compared to Fig.~\ref{fig:mono-abundance-all-sky}). The top panel qualitatively illustrates how clean the metal-poor subsample is: it prominently shows the \emph{Poor Old Heart of the Galaxy} \citep{Rix2022}, without any traces of spurious sample members in the disk that is so dominant in the high-metallicity subsample (bottom panel).}
\label{fig:mono-abundance-all-sky-RGB}
\end{minipage}
\end{figure*}

Since we have an all-sky sample with precise and robust metallicities, it behoves us to make all-sky maps as a function of metallicity to illustrate it.
Fig.~\ref{fig:mono-abundance-all-sky} and Fig.~\ref{fig:mono-abundance-all-sky-RGB} provide all-sky maps of the sample's number density in various metallicity ranges for two samples: first, the complete (unfiltered) sample of  all 174\,922\,161 stars, and second a vetted sample of 17\,558\,141 RGB stars.
The vetted RGB sample was designed to eliminate spurious \MH\ estimates at the expense of sample size, in particular to eliminate sample contamination among the metal-poor stars that result from unrecognized instances of hotter but reddened stars. After some experimentation, we adopted the following selection criteria
illustrated in Fig.~\ref{fig:clean-RGB-cuts}:
\begin{itemize}
\item $\texttt{phot\_g\_mean\_mag} < 16$
\item $\varpi/\sigma_\varpi>4$
\item $\texttt{logg\_xgboost}<3.5$
\item $\texttt{teff\_xgboost}<5200$
\item $M_{W1} > -0.3 - 0.006\cdot(5500-\texttt{teff\_xgboost})$
\item $M_{W1} > -0.01\cdot(5300-\texttt{teff\_xgboost})$
\item $(G - W_2) < 0.2 + 0.77\cdot (G_\textrm{BP} - W_1)$
\end{itemize}
where $M_{W1} = W_1 + 5\cdot \log_{10}(\varpi/100)$.

Only 11\,853 of the 2\,371\,118 of the OBA stars identified by \citet{2022arXiv220605870G} that are in our sample pass these quality cuts, i.e.\ these cuts succeed to eliminate 99.61\% of OBA stars from this sample. Notably, \citet{2022arXiv220605870G} find that their OBA star sample has some contamination from ``halo'' stars (i.e.\ metal-poor stars), which they eliminate kinematically, but cannot eliminate for metal-poor stars with disk-like orbits. Hence, our metal-poor sample may be even purer than the above comparison implies.

For the convenience of the user, this vetted RGB subset is provided as a separate table, as described in Table~\ref{table:vetted-RGB-results}. In this table, we also provide auxiliary information from Gaia about a source's astrometry, photometry, and RVS radial velocities that are available for a substantive fraction of the sample. This provides all the information necessary for the user to compute stellar orbits.

\begin{table}[h!]
\centering
\caption{Table description of 17\,558\,141 vetted RGB results provided online \citep{Andrae2023Zenodo}. We adopt the column names from the Gaia DR3 archive where appropriate. We emphasize that the zero-point correction of \citet{2021A&A...649A...4L} has been applied to the parallaxes in this table.}
\begin{footnotesize}
\label{table:vetted-RGB-results}
\begin{tabular}{c|c}
column name & description \\
\hline
source\_id & GDR3 identifier in ascending order \\
l & Galactic longitude [deg] \\
b &  Galactic lattitude [deg] \\
ra & right ascension [deg] \\
dec & declination [deg] \\
parallax\_corrected & parallax with zero-point correction [mas] \\
parallax\_error & parallax error [mas] \\
pmra & proper motion RA [mas/yr] \\
pmra\_error & error of proper motion RA [mas/yr] \\
pmdec & proper motion DEC [mas/yr] \\
pmdec\_error & error of proper motion DEC [mas/yr] \\
ruwe & astrometric quality flag \\
radial\_velocity & radial velocity [km/s] \\
radial\_velocity\_error & radial velocity error [km/s] \\
phot\_g\_mean\_mag & apparent $G$ magnitude [mag] \\
phot\_bp\_mean\_mag & apparent $G_\textrm{BP}$ magnitude [mag] \\
phot\_rp\_mean\_mag & apparent $G_\textrm{RP}$ magnitude [mag] \\
allwise\_w1 & apparent $W_1$ magnitude [mag] \\
allwise\_w2 & apparent $W_2$ magnitude [mag] \\
mh\_xgboost & XGBoost estimate of \MH \\
teff\_xgboost & XGBoost estimate of \Teff\ [K] \\
logg\_xgboost & XGBoost estimate of \logg \\
in\_training\_sample & membership in training sample \\
\end{tabular}
\end{footnotesize}
\end{table}

The three panels of the all-sky maps for the unfiltered sample in Fig.~\ref{fig:mono-abundance-all-sky} clearly show the imprint of the Gaia scanning law: in particular two ``crescents'' of lower sample density at high latitudes are attributable to too few transits that prevented the publication of XP spectra in Gaia DR3 \citep[c.f.][Fig.~29 therein]{2022arXiv220606143D}. Moreover,  dust extinction in the Galactic plane causes many stars to be dimmed below $G<17.65$: they may be too faint to have XP spectra published or even too faint to be in the Gaia catalog at all. Note that the extinction in the Galactic plane appears most dramatic among the two metal poor bins (top and middle panel), as these have far fewer foreground stars.  The top map in Fig.~\ref{fig:mono-abundance-all-sky} also shows an implausible set of seemingly metal-poor stars near the disk, preferentially in star-forming regions. Most likely, these objects have spuriously low \MH\ estimates, and actually are  reddened OBA stars of presumably higher metallicity that are misinterpreted as metal-poor (see Sect.~\ref{ssec:failure-modes}).

It is these stars that most immediately show that the full unfiltered sample must contain some spurious \MH\ estimates, which motivated our vetted sample of bright giant stars. Comparison of the two top panels of Fig.~\ref{fig:mono-abundance-all-sky} and Fig.~\ref{fig:mono-abundance-all-sky-RGB} shows that these spurious sources are absent in the vetted giant sample.

Apart from these issues, the top maps in Fig.~\ref{fig:mono-abundance-all-sky} and Fig.~\ref{fig:mono-abundance-all-sky-RGB} both clearly show the central concentration of metal-poor stars toward the Galactic center, extensively discussed as the \emph{Poor Old Heart of the Milky Way} in \citet{Rix2022}. Figure~\ref{fig:mono-abundance-all-sky} also shows the two Magellanic clouds, which are missing from Fig.~\ref{fig:mono-abundance-all-sky-RGB} due to the cut on parallax quality of $\frac{\varpi}{\sigma_\varpi}>4$.

\subsection{Potential filtering}
\label{ssec:filtering}

While we have illustrated only two examples of how to vet or filter the overall table of \MH\ estimates, it is clear that there are further limitations of our stellar parameter estimates that may compromise the use of our catalog. Here, we provide some guidance on potential filtering by the user:
\begin{itemize}
\item In \citet{Rix2022}, we used a bright RGB sample defined by teff\_xgboost$<$5300K, logg\_xgboost$<$3.5 and $G_\textrm{BP}<16$. While this is still possible, we point out that our results in this work also hold for main-sequence dwarfs (see Fig.~\ref{fig:internal-x-validation-XGBoost}c). Nevertheless, if the focus is on RGB stars, we recommend to drop or relax the selection on $G_\textrm{BP}$, given that our results are robust towards the faint end (see Fig.~\ref{fig:MH-residuals-vs-apparent-G}).
\item Given that OBA stars are problematic (see Sect.~\ref{ssec:failure-modes}), one could take the golden sample of OBA stars from \citet{2022arXiv220605870G} and remove all known OBA stars from the sample.
\item The Gaia DR3 publication limit of $G=17.65$ for XP spectra was not strict. Rather, XP spectra for 162\,686 QSOs and 26\,500 galaxies were also published down to the survey detection limit. Furthermore, XP spectra of ultra-cool dwarfs were published beyond $G=17.65$. Given our training sample's temperature range of 3107K -- 6867K, ultra-cool dwarfs are not covered. Therefore, the user may consider removing all 129\,997 results for $G>17.65$.
\item The user may want to give special consideration to globular clusters, as those represent regions of high source density where the CCD windows assigned to the XP spectra may begin to overlap, thus compromising the XP spectra and our derived \MH\ estimates. This effect was illustrated for Omega Centauri in Fig.~27 of \citet{2022arXiv220605864C}.
\item When working with \MH\ for stars of the main sequence, we recommend limiting the color range to $0.5<G_\textrm{BP}-G_\textrm{RP}<1.5$ for unbiased results (see Fig.~\ref{fig:MH-distribution-solar-analogues} and Fig.~\ref{fig:Praesepe-MH-vs-colour}). The \Teff\ estimates of stars on the main sequence may be precise over a wider range.
\item In order to prevent invalid extrapolations beyond the training sample, the user can check if a source's colors fall within the training sample, as we did for clusters in Sect.~\ref{ssec:validation-with-clusters} and Fig.~\ref{fig:colour-colour-distribution-application-vs-training-sample}. To this end, our catalog contains a column named \texttt{in\_training\_sample} which is a boolean flag that indicates if a source was part of the training sample.
\end{itemize}

\section{Summary}

We have derived and presented a catalog of data-driven, precise, accurate and robust metallicity estimates \MH\ (as well as \Teff\ and \logg) for 175 million stars from Gaia DR3. These estimates were derived using an externally trained XGBoost algorithm that draws on an extensive set of data features: parallaxes, low-resolution XP spectra, robust synthetic photometry based on those XP spectra, and CatWISE photometry. By construction, the resulting parameters are tied to the stellar parameter scale of the main training set, SDSS DR17 (APOGEE). The entire catalog is published and available online \citep{Andrae2023Zenodo}.

This catalog greatly improves on our earlier catalog in \citet{Rix2022} in several respects: 1) It is all-sky, not restricted to stars towards the Galactic center. 2) It covers much of the stellar color-magnitude plane, not just red giants. 3) It encompasses all stars with XP spectra, not just the bright ones ($G<16$). 4) The \MH\ estimates overcome the $\MH\gtrsim -2.5$ limitation in \citet{Rix2022} by augmenting the main APOGEE DR17 sample by the very and extremely metal poor stars from \citet{2022ApJ...931..147L} in the training of the XGBoost algorithm. 5) It replaces AllWISE \citep{2014yCat.2328....0C} by CatWISE \citep{2021ApJS..253....8M}, thus improving completeness substantially.

For stars within our training sample's temperature range (from 3107K to 6867K), our empirical approach recovers the \MH\ to within an RMS test error of 0.1 (from cross-validation in Fig.~\ref{fig:internal-x-validation-XGBoost}) and an RMS validation error of 0.146 on GALAH DR3 (see Fig.~\ref{fig:external-validation-mh_xgboost}). In particular, our empirical results exhibit the same systematics that our APOGEE-dominated training sample exhibits in comparison to GALAH DR3, i.e.\ our results are perfectly consistent with the known discrepancies between spectroscopic surveys. An independent validation on Solar-analog candidates from \citet{2022arXiv220605870G}, on members of the Praesepe cluster, and on wide binaries from \citet{WideBinaries} confirms typical \MH\ uncertainty of 0.1 with negligible bias for stars on the main sequence in this intermediate temperature regime.  Towards the faint end, \MH\ errors increase moderately, reaching 0.15 at $G\sim 14$, 0.2 at $G\sim 16$, and finally $\sim 0.4$ at $G\sim 17.65$ (see Fig.~\ref{fig:MH-residuals-vs-apparent-G}), but we do not see systematic errors emerge. We suspect that this ``graceful'' degradation is probably caused by increasing noise in the narrow-band photometry synthesized from XP spectra.

We provide the full, unfiltered catalog of all $\sim 175$ million \MH\ estimates without applying any quality cuts (see Table~\ref{table:full-results}), as we had already published a smaller catalog with highly conservative cuts as part of \citet{Rix2022}. The purpose of the current work is to push towards what is maximally possible in \MH\ estimates from XP spectra, to allow further and broader scientific exploitation of the Gaia DR3 data. Consequently, the user is advised to carefully vet each data subset to understand its limitations for each astrophysical application. In particular, all applications that draw on stars with parameters not well represented in the training set, require caution.  We provide some guidance in Sect.~\ref{ssec:filtering}. For user convenience, we also define a vetted sample of 17.5 million RGB stars (see Table~\ref{table:vetted-RGB-results}) with conservative cuts to ensure high data quality. This sample is still much larger than the sample in \citet{Rix2022} which contained only 2 million RGB stars towards the Galactic center. 

However, we emphasize that main-sequence stars from the overall sample presented here can also be used reliably for \MH\ analysis, as long as they are in the color range $0.5<G_\textrm{BP}-G_\textrm{RP}<1.5$, where they achieve typical \MH\ uncertainties between 0.079 for wide binaries (see Sect.~\ref{ssec:wide-binaries}) and 0.1 for Solar analogs (see Fig.~\ref{fig:MH-distribution-solar-analogues}). Additionally, in Appendix~\ref{appendix:ADQL} we provide instructions how to retrieve all Gaia DR3 sources with radial velocity measurements and astrometry and match those with our metallicity catalog in order to facility chemodynamical studies.

This catalog is already being used for various upcoming research projects, e.g.\ on the metallicity gradient in the Large Magellanic Cloud (Andrae et al.\ in prep.), the chemodynamics of the Milky Way disk (Chandra et al.\ in prep.), on stellar rotation in open clusters (Pancino et al.\ in prep.).

Although Gaia~DR3 is only a few months old, our work also allows us to be very optimistic for Gaia DR4: The fact that we can obtain robust and reliable \MH\ estimates even for the faintest stars at the publication limit of XP spectra in Gaia~DR3 (see Fig.~\ref{fig:internal-x-validation-XGBoost}d and Fig.~\ref{fig:MH-residuals-vs-apparent-G}) suggests that useful \MH\ estimates might also be achievable for fainter XP spectra that will be published in Gaia~DR4. In addition, it is likely that the XP spectra themselves will improve substantially from Gaia DR3 to Gaia DR4, due to improved processing and about twice as many observing epochs. All this bodes extremely well for the science potential of the XP spectra that will be published in Gaia~DR4.  

\begin{acknowledgments}
We thank our colleague Morgan Fouesneau for valuable discussions on this work and on the manuscript. In particular, RA thanks Yang Huang for useful background information about LAMOST and SMSS errors and Francois-Xavier Pineau for his valuable help in cross-matching Gaia and CatWISE data.

This work has made use of data from the European Space Agency (ESA) mission {\it Gaia} (\url{https://www.cosmos.esa.int/gaia}), processed by the {\it Gaia} Data Processing and Analysis Consortium (DPAC, \url{https://www.cosmos.esa.int/web/gaia/dpac/consortium}). Funding for the DPAC has been provided by national institutions, in particular the institutions participating in the {\it Gaia} Multilateral Agreement.

Guoshoujing Telescope (the Large Sky Area Multi-Object Fiber Spectroscopic Telescope LAMOST) is a National Major Scientific Project built by the Chinese Academy of Sciences. Funding for the project has been provided by the National Development and Reform Commission. LAMOST is operated and managed by the National Astronomical Observatories, Chinese Academy of Sciences.
\end{acknowledgments}

\software{\texttt{numpy} \citep{Harris2020}, 
\texttt{scipy} \citep{Virtanen2020}, 
\texttt{matplotlib} \citep{Hunter2007}, 
\texttt{astropy} \citep{AstropyCollaboration2013,AstropyCollaboration2018,AstropyCollaboration2022}, \texttt{GaiaXPy} (\url{https://gaia-dpci.github.io/GaiaXPy-website/}), \texttt{gaiadr3\_zeropoint} (\url{https://gitlab.com/icc-ub/public/gaiadr3_zeropoint})
}

\facilities{Gaia, WISE, Sloan, LAMOST, AAT, Skymapper}

\bibliography{bibliography}{}

\begin{thebibliography}{}
\expandafter\ifx\csname natexlab\endcsname\relax\def\natexlab#1{#1}\fi

\bibitem[{{Abdurro'uf} {et~al.}(2022){Abdurro'uf}, {Accetta}, {Aerts}, {Silva
  Aguirre}, {Ahumada}, {Ajgaonkar}, {Filiz Ak}, {Alam}, {Allende Prieto},
  {Almeida}, {Anders}, {Anderson}, {Andrews}, {Anguiano}, {Aquino-Ort{\'\i}z},
  {Arag{\'o}n-Salamanca}, {Argudo-Fern{\'a}ndez}, {Ata}, {Aubert},
  {Avila-Reese}, {Badenes}, {Barb{\'a}}, {Barger}, {Barrera-Ballesteros},
  {Beaton}, {Beers}, {Belfiore}, {Bender}, {Bernardi}, {Bershady}, {Beutler},
  {Bidin}, {Bird}, {Bizyaev}, {Blanc}, {Blanton}, {Boardman}, {Bolton},
  {Boquien}, {Borissova}, {Bovy}, {Brandt}, {Brown}, {Brownstein}, {Brusa},
  {Buchner}, {Bundy}, {Burchett}, {Bureau}, {Burgasser}, {Cabang}, {Campbell},
  {Cappellari}, {Carlberg}, {Wanderley}, {Carrera}, {Cash}, {Chen}, {Chen},
  {Cherinka}, {Chiappini}, {Choi}, {Chojnowski}, {Chung}, {Clerc}, {Cohen},
  {Comerford}, {Comparat}, {da Costa}, {Covey}, {Crane}, {Cruz-Gonzalez},
  {Culhane}, {Cunha}, {Dai}, {Damke}, {Darling}, {Davidson}, {Davies},
  {Dawson}, {De Lee}, {Diamond-Stanic}, {Cano-D{\'\i}az}, {S{\'a}nchez},
  {Donor}, {Duckworth}, {Dwelly}, {Eisenstein}, {Elsworth}, {Emsellem},
  {Eracleous}, {Escoffier}, {Fan}, {Farr}, {Feng}, {Fern{\'a}ndez-Trincado},
  {Feuillet}, {Filipp}, {Fillingham}, {Frinchaboy}, {Fromenteau}, {Galbany},
  {Garc{\'\i}a}, {Garc{\'\i}a-Hern{\'a}ndez}, {Ge}, {Geisler}, {Gelfand},
  {G{\'e}ron}, {Gibson}, {Goddy}, {Godoy-Rivera}, {Grabowski}, {Green},
  {Greener}, {Grier}, {Griffith}, {Guo}, {Guy}, {Hadjara}, {Harding},
  {Hasselquist}, {Hayes}, {Hearty}, {Hern{\'a}ndez}, {Hill}, {Hogg},
  {Holtzman}, {Horta}, {Hsieh}, {Hsu}, {Hsu}, {Huber}, {Huertas-Company},
  {Hutchinson}, {Hwang}, {Ibarra-Medel}, {Chitham}, {Ilha}, {Imig}, {Jaekle},
  {Jayasinghe}, {Ji}, {Johnson}, {Jones}, {J{\"o}nsson}, {Katkov}, {Khalatyan},
  {Kinemuchi}, {Kisku}, {Knapen}, {Kneib}, {Kollmeier}, {Kong}, {Kounkel},
  {Kreckel}, {Krishnarao}, {Lacerna}, {Lane}, {Langgin}, {Lavender}, {Law},
  {Lazarz}, {Leung}, {Leung}, {Lewis}, {Li}, {Li}, {Lian}, {Liang}, {Lin},
  {Lin}, {Lin}, {Lintott}, {Long}, {Longa-Pe{\~n}a}, {L{\'o}pez-Cob{\'a}},
  {Lu}, {Lundgren}, {Luo}, {Mackereth}, {de la Macorra}, {Mahadevan},
  {Majewski}, {Manchado}, {Mandeville}, {Maraston}, {Margalef-Bentabol},
  {Masseron}, {Masters}, {Mathur}, {McDermid}, {Mckay}, {Merloni},
  {Merrifield}, {Meszaros}, {Miglio}, {Di Mille}, {Minniti}, {Minsley},
  {Monachesi}, {Moon}, {Mosser}, {Mulchaey}, {Muna}, {Mu{\~n}oz}, {Myers},
  {Myers}, {Nadathur}, {Nair}, {Nandra}, {Neumann}, {Newman}, {Nidever},
  {Nikakhtar}, {Nitschelm}, {O'Connell}, {Garma-Oehmichen}, {Luan Souza de
  Oliveira}, {Olney}, {Oravetz}, {Ortigoza-Urdaneta}, {Osorio}, {Otter},
  {Pace}, {Padilla}, {Pan}, {Pan}, {Parikh}, {Parker}, {Peirani}, {Pe{\~n}a
  Ram{\'\i}rez}, {Penny}, {Percival}, {Perez-Fournon}, {Pinsonneault},
  {Poidevin}, {Poovelil}, {Price-Whelan}, {B{\'a}rbara de Andrade Queiroz},
  {Raddick}, {Ray}, {Rembold}, {Riddle}, {Riffel}, {Riffel}, {Rix}, {Robin},
  {Rodr{\'\i}guez-Puebla}, {Roman-Lopes}, {Rom{\'a}n-Z{\'u}{\~n}iga}, {Rose},
  {Ross}, {Rossi}, {Rubin}, {Salvato}, {S{\'a}nchez}, {S{\'a}nchez-Gallego},
  {Sanderson}, {Santana Rojas}, {Sarceno}, {Sarmiento}, {Sayres}, {Sazonova},
  {Schaefer}, {Schiavon}, {Schlegel}, {Schneider}, {Schultheis}, {Schwope},
  {Serenelli}, {Serna}, {Shao}, {Shapiro}, {Sharma}, {Shen}, {Shetrone}, {Shu},
  {Simon}, {Skrutskie}, {Smethurst}, {Smith}, {Sobeck}, {Spoo}, {Sprague},
  {Stark}, {Stassun}, {Steinmetz}, {Stello}, {Stone-Martinez},
  {Storchi-Bergmann}, {Stringfellow}, {Stutz}, {Su}, {Taghizadeh-Popp},
  {Talbot}, {Tayar}, {Telles}, {Teske}, {Thakar}, {Theissen}, {Tkachenko},
  {Thomas}, {Tojeiro}, {Hernandez Toledo}, {Troup}, {Trump}, {Trussler},
  {Turner}, {Tuttle}, {Unda-Sanzana}, {V{\'a}zquez-Mata}, {Valentini},
  {Valenzuela}, {Vargas-Gonz{\'a}lez}, {Vargas-Maga{\~n}a}, {Alfaro},
  {Villanova}, {Vincenzo}, {Wake}, {Warfield}, {Washington}, {Weaver},
  {Weijmans}, {Weinberg}, {Weiss}, {Westfall}, {Wild}, {Wilde}, {Wilson},
  {Wilson}, {Wilson}, {Wolf}, {Wood-Vasey}, {Yan}, {Zamora}, {Zasowski},
  {Zhang}, {Zhao}, {Zheng}, {Zheng}, \& {Zhu}}]{SDSS_DR17}
{Abdurro'uf}, {Accetta}, K., {Aerts}, C., {et~al.} 2022, \apjs, 259, 35

\bibitem[{{Almeida} {et~al.}(2023){Almeida}, {Anderson},
  {Argudo-Fern{\'a}ndez}, {Badenes}, {Barger}, {Barrera-Ballesteros}, {Bender},
  {Benitez}, {Besser}, {Bizyaev}, {Blanton}, {Bochanski}, {Bovy}, {Brandt},
  {Brownstein}, {Buchner}, {Bulbul}, {Burchett}, {Cano D{\'\i}az}, {Carlberg},
  {Casey}, {Chandra}, {Cherinka}, {Chiappini}, {Coker}, {Comparat}, {Conroy},
  {Contardo}, {Cortes}, {Covey}, {Crane}, {Cunha}, {Dabbieri}, {Davidson},
  {Davis}, {De Lee}, {M{\'e}ndez Delgado}, {Demasi}, {Di Mille}, {Donor},
  {Dow}, {Dwelly}, {Eracleous}, {Eriksen}, {Fan}, {Farr}, {Frederick}, {Fries},
  {Frinchaboy}, {Gaensicke}, {Ge}, {Gonz{\'a}lez {\'A}vila}, {Grabowski},
  {Grier}, {Guiglion}, {Gupta}, {Hall}, {Hawkins}, {Hayes}, {Hermes},
  {Hern{\'a}ndez-Garc{\'\i}a}, {Hogg}, {Holtzman}, {Ibarra-Medel}, {Ji},
  {Jofre}, {Johnson}, {Jones}, {Kinemuchi}, {Kluge}, {Koekemoer}, {Kollmeier},
  {Kounkel}, {Krishnarao}, {Krumpe}, {Lacerna}, {Jakson Assuncao Lago},
  {Laporte}, {Liu}, {Liu}, {Liu}, {Lopes}, {Macktoobian}, {Malanushenko},
  {Maoz}, {Masseron}, {Masters}, {Matijevic}, {McBride}, {Medan}, {Merloni},
  {Morrison}, {Myers}, {M{\'e}sz{\'a}ros}, {Negrete}, {Nidever}, {Nitschelm},
  {Oravetz}, {Oravetz}, {Pan}, {Peng}, {Pinsonneault}, {Pogge}, {Qiu},
  {Queiroz}, {Ramirez}, {Rix}, {Fern{\'a}ndez Rosso}, {Runnoe}, {Salvato},
  {Sanchez}, {Santana}, {Saydjari}, {Sayres}, {Schlaufman}, {Schneider},
  {Schwope}, {Serna}, {Shen}, {Sobeck}, {Song}, {Souto}, {Spoo}, {Stassun},
  {Steinmetz}, {Straumit}, {Stringfellow}, {S{\'a}nchez-Gallego},
  {Taghizadeh-Popp}, {Tayar}, {Thakar}, {Tissera}, {Tkachenko}, {Hernandez
  Toledo}, {Trakhtenbrot}, {Fernandez Trincado}, {Troup}, {Trump}, {Tuttle},
  {Ulloa}, {Vazquez-Mata}, {Alfaro}, {Villanova}, {Wachter}, {Weijmans},
  {Wheeler}, {Wilson}, {Wojno}, {Wolf}, {Xue}, {Ybarra}, {Zari}, \&
  {Zasowski}}]{SDSS_DR18}
{Almeida}, A., {Anderson}, S.~F., {Argudo-Fern{\'a}ndez}, M., {et~al.} 2023,
  arXiv e-prints, arXiv:2301.07688

\bibitem[{Andrae {et~al.}(2023)Andrae, Rix, \& Chandra}]{Andrae2023Zenodo}
Andrae, R., Rix, H.-W., \& Chandra, V. 2023, {Robust Data-driven Metallicities
  for 120 Million Stars from Gaia XP Spectra}, v.V1.0,  Zenodo,
  doi:10.5281/zenodo.7599789

\bibitem[{{Andrae} {et~al.}(2022){Andrae}, {Fouesneau}, {Sordo},
  {Bailer-Jones}, {Dharmawardena}, {Rybizki}, {De Angeli}, {Lindstr{\o}m},
  {Marshall}, {Drimmel}, {Korn}, {Soubiran}, {Brouillet}, {Casamiquela}, {Rix},
  {Abreu Aramburu}, {{\'A}lvarez}, {Bakker}, {Bellas-Velidis}, {Bijaoui},
  {Brugaletta}, {Burlacu}, {Carballo}, {Chaoul}, {Chiavassa}, {Contursi},
  {Cooper}, {Creevey}, {Dafonte}, {Dapergolas}, {de Laverny}, {Delchambre},
  {Demouchy}, {Edvardsson}, {Fr{\'e}mat}, {Garabato}, {Garc{\'\i}a-Lario},
  {Garc{\'\i}a-Torres}, {Gavel}, {Gomez}, {Gonz{\'a}lez-Santamar{\'\i}a},
  {Hatzidimitriou}, {Heiter}, {Jean-Antoine Piccolo}, {Kontizas}, {Kordopatis},
  {Lanzafame}, {Lebreton}, {Licata}, {Livanou}, {Lobel}, {Lorca}, {Magdaleno
  Romeo}, {Manteiga}, {Marocco}, {Mary}, {Nicolas}, {Ordenovic}, {Pailler},
  {Palicio}, {Pallas-Quintela}, {Panem}, {Pichon}, {Poggio}, {Recio-Blanco},
  {Riclet}, {Robin}, {Santove{\~n}a}, {Sarro}, {Schultheis}, {Segol},
  {Silvelo}, {Slezak}, {Smart}, {S{\"u}veges}, {Th{\'e}venin}, {Torralba
  Elipe}, {Ulla}, {Utrilla}, {Vallenari}, {van Dillen}, {Zhao}, \&
  {Zorec}}]{Andrae22}
{Andrae}, R., {Fouesneau}, M., {Sordo}, R., {et~al.} 2022, arXiv e-prints,
  arXiv:2206.06138

\bibitem[{{Astropy Collaboration} {et~al.}(2013){Astropy Collaboration},
  {Robitaille}, {Tollerud}, {Greenfield}, {Droettboom}, {Bray}, {Aldcroft},
  {Davis}, {Ginsburg}, {Price-Whelan}, {Kerzendorf}, {Conley}, {Crighton},
  {Barbary}, {Muna}, {Ferguson}, {Grollier}, {Parikh}, {Nair}, {Unther},
  {Deil}, {Woillez}, {Conseil}, {Kramer}, {Turner}, {Singer}, {Fox}, {Weaver},
  {Zabalza}, {Edwards}, {Azalee Bostroem}, {Burke}, {Casey}, {Crawford},
  {Dencheva}, {Ely}, {Jenness}, {Labrie}, {Lim}, {Pierfederici}, {Pontzen},
  {Ptak}, {Refsdal}, {Servillat}, \& {Streicher}}]{AstropyCollaboration2013}
{Astropy Collaboration}, {Robitaille}, T.~P., {Tollerud}, E.~J., {et~al.} 2013,
  A\&A, 558, A33

\bibitem[{{Astropy Collaboration} {et~al.}(2018){Astropy Collaboration},
  {Price-Whelan}, {Sip{\H{o}}cz}, {G{\"u}nther}, {Lim}, {Crawford}, {Conseil},
  {Shupe}, {Craig}, {Dencheva}, {Ginsburg}, {VanderPlas}, {Bradley},
  {P{\'e}rez-Su{\'a}rez}, {de Val-Borro}, {Aldcroft}, {Cruz}, {Robitaille},
  {Tollerud}, {Ardelean}, {Babej}, {Bach}, {Bachetti}, {Bakanov}, {Bamford},
  {Barentsen}, {Barmby}, {Baumbach}, {Berry}, {Biscani}, {Boquien}, {Bostroem},
  {Bouma}, {Brammer}, {Bray}, {Breytenbach}, {Buddelmeijer}, {Burke},
  {Calderone}, {Cano Rodr{\'\i}guez}, {Cara}, {Cardoso}, {Cheedella}, {Copin},
  {Corrales}, {Crichton}, {D'Avella}, {Deil}, {Depagne}, {Dietrich}, {Donath},
  {Droettboom}, {Earl}, {Erben}, {Fabbro}, {Ferreira}, {Finethy}, {Fox},
  {Garrison}, {Gibbons}, {Goldstein}, {Gommers}, {Greco}, {Greenfield},
  {Groener}, {Grollier}, {Hagen}, {Hirst}, {Homeier}, {Horton}, {Hosseinzadeh},
  {Hu}, {Hunkeler}, {Ivezi{\'c}}, {Jain}, {Jenness}, {Kanarek}, {Kendrew},
  {Kern}, {Kerzendorf}, {Khvalko}, {King}, {Kirkby}, {Kulkarni}, {Kumar},
  {Lee}, {Lenz}, {Littlefair}, {Ma}, {Macleod}, {Mastropietro}, {McCully},
  {Montagnac}, {Morris}, {Mueller}, {Mumford}, {Muna}, {Murphy}, {Nelson},
  {Nguyen}, {Ninan}, {N{\"o}the}, {Ogaz}, {Oh}, {Parejko}, {Parley}, {Pascual},
  {Patil}, {Patil}, {Plunkett}, {Prochaska}, {Rastogi}, {Reddy Janga},
  {Sabater}, {Sakurikar}, {Seifert}, {Sherbert}, {Sherwood-Taylor}, {Shih},
  {Sick}, {Silbiger}, {Singanamalla}, {Singer}, {Sladen}, {Sooley},
  {Sornarajah}, {Streicher}, {Teuben}, {Thomas}, {Tremblay}, {Turner},
  {Terr{\'o}n}, {van Kerkwijk}, {de la Vega}, {Watkins}, {Weaver}, {Whitmore},
  {Woillez}, {Zabalza}, \& {Astropy Contributors}}]{AstropyCollaboration2018}
{Astropy Collaboration}, {Price-Whelan}, A.~M., {Sip{\H{o}}cz}, B.~M., {et~al.}
  2018, AJ, 156, 123

\bibitem[{{Astropy Collaboration} {et~al.}(2022){Astropy Collaboration},
  {Price-Whelan}, {Lim}, {Earl}, {Starkman}, {Bradley}, {Shupe}, {Patil},
  {Corrales}, {Brasseur}, {N{\"o}the}, {Donath}, {Tollerud}, {Morris},
  {Ginsburg}, {Vaher}, {Weaver}, {Tocknell}, {Jamieson}, {van Kerkwijk},
  {Robitaille}, {Merry}, {Bachetti}, {G{\"u}nther}, {Aldcroft},
  {Alvarado-Montes}, {Archibald}, {B{\'o}di}, {Bapat}, {Barentsen},
  {Baz{\'a}n}, {Biswas}, {Boquien}, {Burke}, {Cara}, {Cara}, {Conroy},
  {Conseil}, {Craig}, {Cross}, {Cruz}, {D'Eugenio}, {Dencheva}, {Devillepoix},
  {Dietrich}, {Eigenbrot}, {Erben}, {Ferreira}, {Foreman-Mackey}, {Fox},
  {Freij}, {Garg}, {Geda}, {Glattly}, {Gondhalekar}, {Gordon}, {Grant},
  {Greenfield}, {Groener}, {Guest}, {Gurovich}, {Handberg}, {Hart},
  {Hatfield-Dodds}, {Homeier}, {Hosseinzadeh}, {Jenness}, {Jones}, {Joseph},
  {Kalmbach}, {Karamehmetoglu}, {Ka{\l}uszy{\'n}ski}, {Kelley}, {Kern},
  {Kerzendorf}, {Koch}, {Kulumani}, {Lee}, {Ly}, {Ma}, {MacBride}, {Maljaars},
  {Muna}, {Murphy}, {Norman}, {O'Steen}, {Oman}, {Pacifici}, {Pascual},
  {Pascual-Granado}, {Patil}, {Perren}, {Pickering}, {Rastogi}, {Roulston},
  {Ryan}, {Rykoff}, {Sabater}, {Sakurikar}, {Salgado}, {Sanghi}, {Saunders},
  {Savchenko}, {Schwardt}, {Seifert-Eckert}, {Shih}, {Jain}, {Shukla}, {Sick},
  {Simpson}, {Singanamalla}, {Singer}, {Singhal}, {Sinha}, {Sip{\H{o}}cz},
  {Spitler}, {Stansby}, {Streicher}, {{\v{S}}umak}, {Swinbank}, {Taranu},
  {Tewary}, {Tremblay}, {de Val-Borro}, {Van Kooten}, {Vasovi{\'c}}, {Verma},
  {de Miranda Cardoso}, {Williams}, {Wilson}, {Winkel}, {Wood-Vasey}, {Xue},
  {Yoachim}, {Zhang}, {Zonca}, \& {Astropy Project
  Contributors}}]{AstropyCollaboration2022}
{Astropy Collaboration}, {Price-Whelan}, A.~M., {Lim}, P.~L., {et~al.} 2022,
  ApJ, 935, 167

\bibitem[{{Belokurov} {et~al.}(2023){Belokurov}, {Vasiliev}, {Deason},
  {Koposov}, {Fattahi}, {Dillamore}, {Davies}, \& {Grand}}]{Belokurov2023}
{Belokurov}, V., {Vasiliev}, E., {Deason}, A.~J., {et~al.} 2023, MNRAS, 518,
  6200

\bibitem[{{Buder} {et~al.}(2021){Buder}, {Sharma}, {Kos}, {Amarsi},
  {Nordlander}, {Lind}, {Martell}, {Asplund}, {Bland-Hawthorn}, {Casey}, {de
  Silva}, {D'Orazi}, {Freeman}, {Hayden}, {Lewis}, {Lin}, {Schlesinger},
  {Simpson}, {Stello}, {Zucker}, {Zwitter}, {Beeson}, {Buck}, {Casagrande},
  {Clark}, {{\v{C}}otar}, {da Costa}, {de Grijs}, {Feuillet}, {Horner},
  {Kafle}, {Khanna}, {Kobayashi}, {Liu}, {Montet}, {Nandakumar}, {Nataf},
  {Ness}, {Spina}, {Tepper-Garc{\'\i}a}, {Ting}, {Traven},
  {Vogrin{\v{c}}i{\v{c}}}, {Wittenmyer}, {Wyse}, {{\v{Z}}erjal}, \& {Galah
  Collaboration}}]{GALAH_DR3}
{Buder}, S., {Sharma}, S., {Kos}, J., {et~al.} 2021, \mnras, 506, 150

\bibitem[{{Carrasco} {et~al.}(2021){Carrasco}, {Weiler}, {Jordi}, {Fabricius},
  {De Angeli}, {Evans}, {van Leeuwen}, {Riello}, \&
  {Montegriffo}}]{2021A&A...652A..86C}
{Carrasco}, J.~M., {Weiler}, M., {Jordi}, C., {et~al.} 2021, \aap, 652, A86

\bibitem[{{Chandra} {et~al.}(2023){Chandra}, {Naidu}, {Conroy}, {Ji}, {Rix},
  {Bonaca}, {Cargile}, {Han}, {Johnson}, {Ting}, {Woody}, \&
  {Zaritsky}}]{Chandra2023}
{Chandra}, V., {Naidu}, R.~P., {Conroy}, C., {et~al.} 2023, arXiv,
  arXiv:2212.00806

\bibitem[{Chen \& Guestrin(2016)}]{Chen:2016:XST:2939672.2939785}
Chen, T., \& Guestrin, C. 2016, in Proceedings of the 22nd ACM SIGKDD
  International Conference on Knowledge Discovery and Data Mining, KDD '16 (New
  York, NY, USA: ACM), 785--794

\bibitem[{{Chiti} {et~al.}(2021){Chiti}, {Frebel}, {Mardini}, {Daniel}, {Ou},
  \& {Uvarova}}]{2021ApJS..254...31C}
{Chiti}, A., {Frebel}, A., {Mardini}, M.~K., {et~al.} 2021, \apjs, 254, 31

\bibitem[{{Creevey} {et~al.}(2022){Creevey}, {Sordo}, {Pailler}, {Fr{\'e}mat},
  {Heiter}, {Th{\'e}venin}, {Andrae}, {Fouesneau}, {Lobel}, {Bailer-Jones},
  {Garabato}, {Bellas-Velidis}, {Brugaletta}, {Lorca}, {Ordenovic}, {Palicio},
  {Sarro}, {Delchambre}, {Drimmel}, {Rybizki}, {Torralba Elipe}, {Korn},
  {Recio-Blanco}, {Schultheis}, {De Angeli}, {Montegriffo}, {Abreu Aramburu},
  {Accart}, {{\'A}lvarez}, {Bakker}, {Brouillet}, {Burlacu}, {Carballo},
  {Casamiquela}, {Chiavassa}, {Contursi}, {Cooper}, {Dafonte}, {Dapergolas},
  {de Laverny}, {Dharmawardena}, {Edvardsson}, {Le Fustec},
  {Garc{\'\i}a-Lario}, {Garc{\'\i}a-Torres}, {Gomez},
  {Gonz{\'a}lez-Santamar{\'\i}a}, {Hatzidimitriou}, {Jean-Antoine Piccolo},
  {Kontizas}, {Kordopatis}, {Lanzafame}, {Lebreton}, {Licata}, {Lindstr{\o}m},
  {Livanou}, {Magdaleno Romeo}, {Manteiga}, {Marocco}, {Marshall}, {Mary},
  {Nicolas}, {Pallas-Quintela}, {Panem}, {Pichon}, {Poggio}, {Riclet}, {Robin},
  {Santove{\~n}a}, {Silvelo}, {Slezak}, {Smart}, {Soubiran}, {S{\"u}veges},
  {Ulla}, {Utrilla}, {Vallenari}, {Zhao}, {Zorec}, {Barrado}, {Bijaoui},
  {Bouret}, {Blomme}, {Brott}, {Cassisi}, {Kochukhov}, {Martayan}, {Shulyak},
  \& {Silvester}}]{2022arXiv220605864C}
{Creevey}, O.~L., {Sordo}, R., {Pailler}, F., {et~al.} 2022, arXiv e-prints,
  arXiv:2206.05864

\bibitem[{{Cui} {et~al.}(2012){Cui}, {Zhao}, {Chu}, {Li}, {Li}, {Zhang}, {Su},
  {Yao}, {Wang}, {Xing}, {Li}, {Zhu}, {Wang}, {Gu}, {Luo}, {Xu}, {Zhang},
  {Liu}, {Zhang}, {Yang}, {Cao}, {Chen}, {Chen}, {Chen}, {Chen}, {Chu}, {Feng},
  {Gong}, {Hou}, {Hu}, {Hu}, {Hu}, {Jia}, {Jiang}, {Jiang}, {Jiang}, {Jin},
  {Li}, {Li}, {Li}, {Liu}, {Liu}, {Lu}, {Mao}, {Men}, {Qi}, {Qi}, {Shi},
  {Tang}, {Tao}, {Wang}, {Wang}, {Wang}, {Wang}, {Wang}, {Wang}, {Wang},
  {Wang}, {Wang}, {Wang}, {Wang}, {Wang}, {Xu}, {Xu}, {Yang}, {Yu}, {Yuan},
  {Yuan}, {Zhai}, {Zhang}, {Zhang}, {Zhang}, {Zhao}, {Zhou}, {Zhou}, {Zhu}, \&
  {Zou}}]{Cui2012}
{Cui}, X.-Q., {Zhao}, Y.-H., {Chu}, Y.-Q., {et~al.} 2012, RAA, 12, 1197

\bibitem[{{Cutri} {et~al.}(2021){Cutri}, {Wright}, {Conrow}, {Fowler},
  {Eisenhardt}, {Grillmair}, {Kirkpatrick}, {Masci}, {McCallon}, {Wheelock},
  {Fajardo-Acosta}, {Yan}, {Benford}, {Harbut}, {Jarrett}, {Lake}, {Leisawitz},
  {Ressler}, {Stanford}, {Tsai}, {Liu}, {Helou}, {Mainzer}, {Gettngs},
  {Gonzalez}, {Hoffman}, {Marsh}, {Padgett}, {Skrutskie}, {Beck}, {Papin}, \&
  {Wittman}}]{2014yCat.2328....0C}
{Cutri}, R.~M., {Wright}, E.~L., {Conrow}, T., {et~al.} 2021, VizieR Online
  Data Catalog, II/328

\bibitem[{{Dalton} {et~al.}(2012){Dalton}, {Trager}, {Abrams}, {Carter},
  {Bonifacio}, {Aguerri}, {MacIntosh}, {Evans}, {Lewis}, {Navarro}, {Agocs},
  {Dee}, {Rousset}, {Tosh}, {Middleton}, {Pragt}, {Terrett}, {Brock}, {Benn},
  {Verheijen}, {Cano Infantes}, {Bevil}, {Steele}, {Mottram}, {Bates},
  {Gribbin}, {Rey}, {Rodriguez}, {Delgado}, {Guinouard}, {Walton}, {Irwin},
  {Jagourel}, {Stuik}, {Gerlofsma}, {Roelfsma}, {Skillen}, {Ridings},
  {Balcells}, {Daban}, {Gouvret}, {Venema}, \& {Girard}}]{Dalton2012}
{Dalton}, G., {Trager}, S.~C., {Abrams}, D.~C., {et~al.} 2012, in Society of
  Photo-Optical Instrumentation Engineers (SPIE) Conference Series, Vol. 8446,
  Ground-based and Airborne Instrumentation for Astronomy IV, ed. I.~S.
  {McLean}, S.~K. {Ramsay}, \& H.~{Takami}, 84460P

\bibitem[{{De Angeli} {et~al.}(2022){De Angeli}, {Weiler}, {Montegriffo},
  {Evans}, {Riello}, {Andrae}, {Carrasco}, {Busso}, {Burgess}, {Cacciari},
  {Davidson}, {Harrison}, {Hodgkin}, {Jordi}, {Osborne}, {Pancino},
  {Altavilla}, {Barstow}, {Bailer-Jones}, {Bellazzini}, {Brown}, {Castellani},
  {Cowell}, {Delchambre}, {De Luise}, {Diener}, {Fabricius}, {Fouesneau},
  {Fremat}, {Gilmore}, {Giuffrida}, {Hambly}, {Hidalgo}, {Holland},
  {Kostrzewa-Rutkowska}, {van Leeuwen}, {Lobel}, {Marinoni}, {Miller},
  {Pagani}, {Palaversa}, {Piersimoni}, {Pulone}, {Ragaini}, {Rainer},
  {Richards}, {Rixon}, {Ruz-Mieres}, {Sanna}, {Sarro}, {Rowell}, {Sordo},
  {Walton}, \& {Yoldas}}]{2022arXiv220606143D}
{De Angeli}, F., {Weiler}, M., {Montegriffo}, P., {et~al.} 2022, arXiv
  e-prints, arXiv:2206.06143

\bibitem[{{de Jong} {et~al.}(2019){de Jong}, {Agertz}, {Berbel}, {Aird},
  {Alexander}, {Amarsi}, {Anders}, {Andrae}, {Ansarinejad}, {Ansorge}, \&
  et~al.}]{deJong2019}
{de Jong}, R.~S., {Agertz}, O., {Berbel}, A.~A., {et~al.} 2019, Msngr, 175, 3

\bibitem[{{De Silva} {et~al.}(2015){De Silva}, {Freeman}, {Bland-Hawthorn},
  {Martell}, {de Boer}, {Asplund}, {Keller}, {Sharma}, {Zucker}, {Zwitter},
  {Anguiano}, {Bacigalupo}, {Bayliss}, {Beavis}, {Bergemann}, {Campbell},
  {Cannon}, {Carollo}, {Casagrande}, {Casey}, {Da Costa}, {D'Orazi}, {Dotter},
  {Duong}, {Heger}, {Ireland}, {Kafle}, {Kos}, {Lattanzio}, {Lewis}, {Lin},
  {Lind}, {Munari}, {Nataf}, {O'Toole}, {Parker}, {Reid}, {Schlesinger},
  {Sheinis}, {Simpson}, {Stello}, {Ting}, {Traven}, {Watson}, {Wittenmyer},
  {Yong}, \& {{\v{Z}}erjal}}]{DeSilva2015}
{De Silva}, G.~M., {Freeman}, K.~C., {Bland-Hawthorn}, J., {et~al.} 2015,
  MNRAS, 449, 2604

\bibitem[{{Dekel} \& {Silk}(1986)}]{Dekel1986}
{Dekel}, A., \& {Silk}, J. 1986, ApJ, 303, 39

\bibitem[{{El-Badry} {et~al.}(2021){El-Badry}, {Rix}, \&
  {Heintz}}]{WideBinaries}
{El-Badry}, K., {Rix}, H.-W., \& {Heintz}, T.~M. 2021, \mnras, 506, 2269

\bibitem[{{Gaia Collaboration} {et~al.}(2018){Gaia Collaboration}, {Babusiaux},
  {van Leeuwen}, {Barstow}, {Jordi}, {Vallenari}, {Bossini}, {Bressan},
  {Cantat-Gaudin}, {van Leeuwen}, {Brown}, {Prusti}, {de Bruijne},
  {Bailer-Jones}, {Biermann}, {Evans}, {Eyer}, {Jansen}, {Klioner}, {Lammers},
  {Lindegren}, {Luri}, {Mignard}, {Panem}, {Pourbaix}, {Randich}, {Sartoretti},
  {Siddiqui}, {Soubiran}, {Walton}, {Arenou}, {Bastian}, {Cropper}, {Drimmel},
  {Katz}, {Lattanzi}, {Bakker}, {Cacciari}, {Casta{\~n}eda}, {Chaoul}, {Cheek},
  {De Angeli}, {Fabricius}, {Guerra}, {Holl}, {Masana}, {Messineo}, {Mowlavi},
  {Nienartowicz}, {Panuzzo}, {Portell}, {Riello}, {Seabroke}, {Tanga},
  {Th{\'e}venin}, {Gracia-Abril}, {Comoretto}, {Garcia-Reinaldos}, {Teyssier},
  {Altmann}, {Andrae}, {Audard}, {Bellas-Velidis}, {Benson}, {Berthier},
  {Blomme}, {Burgess}, {Busso}, {Carry}, {Cellino}, {Clementini}, {Clotet},
  {Creevey}, {Davidson}, {De Ridder}, {Delchambre}, {Dell'Oro}, {Ducourant},
  {Fern{\'a}ndez-Hern{\'a}ndez}, {Fouesneau}, {Fr{\'e}mat}, {Galluccio},
  {Garc{\'\i}a-Torres}, {Gonz{\'a}lez-N{\'u}{\~n}ez}, {Gonz{\'a}lez-Vidal},
  {Gosset}, {Guy}, {Halbwachs}, {Hambly}, {Harrison}, {Hern{\'a}ndez},
  {Hestroffer}, {Hodgkin}, {Hutton}, {Jasniewicz}, {Jean-Antoine-Piccolo},
  {Jordan}, {Korn}, {Krone-Martins}, {Lanzafame}, {Lebzelter}, {L{\"o}ffler},
  {Manteiga}, {Marrese}, {Mart{\'\i}n-Fleitas}, {Moitinho}, {Mora}, {Muinonen},
  {Osinde}, {Pancino}, {Pauwels}, {Petit}, {Recio-Blanco}, {Richards},
  {Rimoldini}, {Robin}, {Sarro}, {Siopis}, {Smith}, {Sozzetti}, {S{\"u}veges},
  {Torra}, {van Reeven}, {Abbas}, {Abreu Aramburu}, {Accart}, {Aerts},
  {Altavilla}, {{\'A}lvarez}, {Alvarez}, {Alves}, {Anderson}, {Andrei},
  {Anglada Varela}, {Antiche}, {Antoja}, {Arcay}, {Astraatmadja}, {Bach},
  {Baker}, {Balaguer-N{\'u}{\~n}ez}, {Balm}, {Barache}, {Barata}, {Barbato},
  {Barblan}, {Barklem}, {Barrado}, {Barros}, {Bartholom{\'e} Mu{\~n}oz},
  {Bassilana}, {Becciani}, {Bellazzini}, {Berihuete}, {Bertone}, {Bianchi},
  {Bienaym{\'e}}, {Blanco-Cuaresma}, {Boch}, {Boeche}, {Bombrun}, {Borrachero},
  {Bouquillon}, {Bourda}, {Bragaglia}, {Bramante}, {Breddels}, {Brouillet},
  {Br{\"u}semeister}, {Brugaletta}, {Bucciarelli}, {Burlacu}, {Busonero},
  {Butkevich}, {Buzzi}, {Caffau}, {Cancelliere}, {Cannizzaro}, {Carballo},
  {Carlucci}, {Carrasco}, {Casamiquela}, {Castellani}, {Castro-Ginard},
  {Charlot}, {Chemin}, {Chiavassa}, {Cocozza}, {Costigan}, {Cowell}, {Crifo},
  {Crosta}, {Crowley}, {Cuypers}, {Dafonte}, {Damerdji}, {Dapergolas}, {David},
  {David}, {de Laverny}, {De Luise}, {De March}, {de Martino}, {de Souza}, {de
  Torres}, {Debosscher}, {del Pozo}, {Delbo}, {Delgado}, {Delgado}, {Diakite},
  {Diener}, {Distefano}, {Dolding}, {Drazinos}, {Dur{\'a}n}, {Edvardsson},
  {Enke}, {Eriksson}, {Esquej}, {Eynard Bontemps}, {Fabre}, {Fabrizio},
  {Faigler}, {Falc{\~a}o}, {Farr{\`a}s Casas}, {Federici}, {Fedorets},
  {Fernique}, {Figueras}, {Filippi}, {Findeisen}, {Fonti}, {Fraile}, {Fraser},
  {Fr{\'e}zouls}, {Gai}, {Galleti}, {Garabato}, {Garc{\'\i}a-Sedano},
  {Garofalo}, {Garralda}, {Gavel}, {Gavras}, {Gerssen}, {Geyer}, {Giacobbe},
  {Gilmore}, {Girona}, {Giuffrida}, {Glass}, {Gomes}, {Granvik}, {Gueguen},
  {Guerrier}, {Guiraud}, {Guti{\'e}}, {Haigron}, {Hatzidimitriou}, {Hauser},
  {Haywood}, {Heiter}, {Helmi}, {Heu}, {Hilger}, {Hobbs}, {Hofmann}, {Holland},
  {Huckle}, {Hypki}, {Icardi}, {Jan{\ss}en}, {Jevardat de Fombelle}, {Jonker},
  {Juh{\'a}sz}, {Julbe}, {Karampelas}, {Kewley}, {Klar}, {Kochoska}, {Kohley},
  {Kolenberg}, {Kontizas}, {Kontizas}, {Koposov}, {Kordopatis},
  {Kostrzewa-Rutkowska}, {Koubsky}, {Lambert}, {Lanza}, {Lasne}, {Lavigne}, {Le
  Fustec}, {Le Poncin-Lafitte}, {Lebreton}, {Leccia}, {Leclerc},
  {Lecoeur-Taibi}, {Lenhardt}, {Leroux}, {Liao}, {Licata}, {Lindstr{\o}m},
  {Lister}, {Livanou}, {Lobel}, {L{\'o}pez}, {Managau}, {Mann}, {Mantelet},
  {Marchal}, {Marchant}, {Marconi}, {Marinoni}, {Marschalk{\'o}}, {Marshall},
  {Martino}, {Marton}, {Mary}, {Massari}, {Matijevi{\v{c}}}, {Mazeh},
  {McMillan}, {Messina}, {Michalik}, {Millar}, {Molina}, {Molinaro},
  {Moln{\'a}r}, {Montegriffo}, {Mor}, {Morbidelli}, {Morel}, {Morris},
  {Mulone}, {Muraveva}, {Musella}, {Nelemans}, {Nicastro}, {Noval},
  {O'Mullane}, {Ord{\'e}novic}, {Ord{\'o}{\~n}ez-Blanco}, {Osborne}, {Pagani},
  {Pagano}, {Pailler}, {Palacin}, {Palaversa}, {Panahi}, {Pawlak},
  {Piersimoni}, {Pineau}, {Plachy}, {Plum}, {Poggio}, {Poujoulet},
  {Pr{\v{s}}a}, {Pulone}, {Racero}, {Ragaini}, {Rambaux}, {Ramos-Lerate},
  {Regibo}, {Reyl{\'e}}, {Riclet}, {Ripepi}, {Riva}, {Rivard}, {Rixon},
  {Roegiers}, {Roelens}, {Romero-G{\'o}mez}, {Rowell}, {Royer}, {Ruiz-Dern},
  {Sadowski}, {Sagrist{\`a} Sell{\'e}s}, {Sahlmann}, {Salgado}, {Salguero},
  {Sanna}, {Santana-Ros}, {Sarasso}, {Savietto}, {Schultheis}, {Sciacca},
  {Segol}, {Segovia}, {S{\'e}gransan}, {Shih}, {Siltala}, {Silva}, {Smart},
  {Smith}, {Solano}, {Solitro}, {Sordo}, {Soria Nieto}, {Souchay}, {Spagna},
  {Spoto}, {Stampa}, {Steele}, {Steidelm{\"u}ller}, {Stephenson}, {Stoev},
  {Suess}, {Surdej}, {Szabados}, {Szegedi-Elek}, {Tapiador}, {Taris}, {Tauran},
  {Taylor}, {Teixeira}, {Terrett}, {Teyssandier}, {Thuillot}, {Titarenko},
  {Torra Clotet}, {Turon}, {Ulla}, {Utrilla}, {Uzzi}, {Vaillant}, {Valentini},
  {Valette}, {van Elteren}, {Van Hemelryck}, {Vaschetto}, {Vecchiato},
  {Veljanoski}, {Viala}, {Vicente}, {Vogt}, {von Essen}, {Voss}, {Votruba},
  {Voutsinas}, {Walmsley}, {Weiler}, {Wertz}, {Wevers}, {Wyrzykowski},
  {Yoldas}, {{\v{Z}}erjal}, {Ziaeepour}, {Zorec}, {Zschocke}, {Zucker},
  {Zurbach}, \& {Zwitter}}]{2018A&A...616A..10G}
{Gaia Collaboration}, {Babusiaux}, C., {van Leeuwen}, F., {et~al.} 2018, \aap,
  616, A10

\bibitem[{{Gaia Collaboration} {et~al.}(2022{\natexlab{a}}){Gaia
  Collaboration}, {Creevey}, {Sarro}, {Lobel}, {Pancino}, {Andrae}, {Smart},
  {Clementini}, {Heiter}, {Korn}, {Fouesneau}, {Fr{\'e}mat}, {De Angeli},
  {Vallenari}, {Harrison}, {Th{\'e}venin}, {Reyl{\'e}}, {Sordo}, {Garofalo},
  {Brown}, {Eyer}, {Prusti}, {de Bruijne}, {Arenou}, {Babusiaux}, {Biermann},
  {Ducourant}, {Evans}, {Guerra}, {Hutton}, {Jordi}, {Klioner}, {Lammers},
  {Lindegren}, {Luri}, {Mignard}, {Panem}, {Pourbaix}, {Randich}, {Sartoretti},
  {Soubiran}, {Tanga}, {Walton}, {Bailer-Jones}, {Bastian}, {Drimmel},
  {Jansen}, {Katz}, {Lattanzi}, {van Leeuwen}, {Bakker}, {Cacciari},
  {Casta{\~n}eda}, {Fabricius}, {Galluccio}, {Guerrier}, {Masana}, {Messineo},
  {Mowlavi}, {Nicolas}, {Nienartowicz}, {Pailler}, {Panuzzo}, {Riclet}, {Roux},
  {Seabroke}, {Gracia-Abril}, {Portell}, {Teyssier}, {Altmann}, {Audard},
  {Bellas-Velidis}, {Benson}, {Berthier}, {Blomme}, {Burgess}, {Busonero},
  {Busso}, {C{\'a}novas}, {Carry}, {Cellino}, {Cheek}, {Damerdji}, {Davidson},
  {de Teodoro}, {Nu{\~n}ez Campos}, {Delchambre}, {Dell'Oro}, {Esquej},
  {Fern{\'a}ndez-Hern{\'a}ndez}, {Fraile}, {Garabato}, {Garc{\'\i}a-Lario},
  {Gosset}, {Haigron}, {Halbwachs}, {Hambly}, {Hern{\'a}ndez}, {Hestroffer},
  {Hodgkin}, {Holl}, {Jan{\ss}en}, {Jevardat de Fombelle}, {Jordan},
  {Krone-Martins}, {Lanzafame}, {L{\"o}ffler}, {Marchal}, {Marrese},
  {Moitinho}, {Muinonen}, {Osborne}, {Pauwels}, {Recio-Blanco}, {Riello},
  {Rimoldini}, {Roegiers}, {Rybizki}, {Siopis}, {Smith}, {Sozzetti}, {Utrilla},
  {van Leeuwen}, {Abbas}, {{\'A}brah{\'a}m}, {Abreu Aramburu}, {Aerts},
  {Aguado}, {Ajaj}, {Aldea-Montero}, {Altavilla}, {{\'A}lvarez}, {Alves},
  {Anders}, {Anderson}, {Anglada Varela}, {Antoja}, {Baines}, {Baker},
  {Balaguer-N{\'u}{\~n}ez}, {Balbinot}, {Balog}, {Barache}, {Barbato},
  {Barros}, {Barstow}, {Bartolom{\'e}}, {Bassilana}, {Bauchet}, {Becciani},
  {Bellazzini}, {Berihuete}, {Bernet}, {Bertone}, {Bianchi}, {Binnenfeld},
  {Blanco-Cuaresma}, {Boch}, {Bombrun}, {Bossini}, {Bouquillon}, {Bragaglia},
  {Bramante}, {Breedt}, {Bressan}, {Brouillet}, {Brugaletta}, {Bucciarelli},
  {Burlacu}, {Butkevich}, {Buzzi}, {Caffau}, {Cancelliere}, {Cantat-Gaudin},
  {Carballo}, {Carlucci}, {Carnerero}, {Carrasco}, {Casamiquela}, {Castellani},
  {Castro-Ginard}, {Chaoul}, {Charlot}, {Chemin}, {Chiaramida}, {Chiavassa},
  {Chornay}, {Comoretto}, {Contursi}, {Cooper}, {Cornez}, {Cowell}, {Crifo},
  {Cropper}, {Crosta}, {Crowley}, {Dafonte}, {Dapergolas}, {David}, {de
  Laverny}, {De Luise}, {De March}, {De Ridder}, {de Souza}, {de Torres}, {del
  Peloso}, {del Pozo}, {Delbo}, {Delgado}, {Delisle}, {Demouchy},
  {Dharmawardena}, {Di Matteo}, {Diakite}, {Diener}, {Distefano}, {Dolding},
  {Enke}, {Fabre}, {Fabrizio}, {Faigler}, {Fedorets}, {Fernique}, {Figueras},
  {Fournier}, {Fouron}, {Fragkoudi}, {Gai}, {Garcia-Gutierrez},
  {Garcia-Reinaldos}, {Garc{\'\i}a-Torres}, {Gavel}, {Gavras}, {Gerlach},
  {Geyer}, {Giacobbe}, {Gilmore}, {Girona}, {Giuffrida}, {Gomel}, {Gomez},
  {Gonz{\'a}lez-N{\'u}{\~n}ez}, {Gonz{\'a}lez-Santamar{\'\i}a},
  {Gonz{\'a}lez-Vidal}, {Granvik}, {Guillout}, {Guiraud},
  {Guti{\'e}rrez-S{\'a}nchez}, {Guy}, {Hatzidimitriou}, {Hauser}, {Haywood},
  {Helmer}, {Helmi}, {Sarmiento}, {Hidalgo}, {H{\l}adczuk}, {Hobbs}, {Holland},
  {Huckle}, {Jardine}, {Jasniewicz}, {Jean-Antoine Piccolo},
  {Jim{\'e}nez-Arranz}, {Juaristi Campillo}, {Julbe}, {Karbevska}, {Kervella},
  {Khanna}, {Kordopatis}, {K{\'o}sp{\'a}l}, {Kostrzewa-Rutkowska},
  {Kruszy{\'n}ska}, {Kun}, {Laizeau}, {Lambert}, {Lanza}, {Lasne}, {Le
  Campion}, {Lebreton}, {Lebzelter}, {Leccia}, {Leclerc}, {Lecoeur-Taibi},
  {Liao}, {Licata}, {Lindstr{\o}m}, {Lister}, {Livanou}, {Lorca}, {Loup},
  {Madrero Pardo}, {Magdaleno Romeo}, {Managau}, {Mann}, {Manteiga},
  {Marchant}, {Marconi}, {Marcos}, {Marcos Santos}, {Mar{\'\i}n Pina},
  {Marinoni}, {Marocco}, {Marshall}, {Polo}, {Mart{\'\i}n-Fleitas}, {Marton},
  {Mary}, {Masip}, {Massari}, {Mastrobuono-Battisti}, {Mazeh}, {McMillan},
  {Messina}, {Michalik}, {Millar}, {Mints}, {Molina}, {Molinaro}, {Moln{\'a}r},
  {Monari}, {Mongui{\'o}}, {Montegriffo}, {Montero}, {Mor}, {Mora},
  {Morbidelli}, {Morel}, {Morris}, {Muraveva}, {Murphy}, {Musella}, {Nagy},
  {Noval}, {Oca{\~n}a}, {Ogden}, {Ordenovic}, {Osinde}, {Pagani}, {Pagano},
  {Palaversa}, {Palicio}, {Pallas-Quintela}, {Panahi}, {Payne-Wardenaar},
  {Pe{\~n}alosa Esteller}, {Penttil{\"a}}, {Pichon}, {Piersimoni}, {Pineau},
  {Plachy}, {Plum}, {Poggio}, {Pr{\v{s}}a}, {Pulone}, {Racero}, {Ragaini},
  {Rainer}, {Raiteri}, {Ramos}, {Ramos-Lerate}, {Re Fiorentin}, {Regibo},
  {Richards}, {Rios Diaz}, {Ripepi}, {Riva}, {Rix}, {Rixon}, {Robichon},
  {Robin}, {Robin}, {Roelens}, {Rogues}, {Rohrbasser}, {Romero-G{\'o}mez},
  {Rowell}, {Royer}, {Ruz Mieres}, {Rybicki}, {Sadowski}, {S{\'a}ez
  N{\'u}{\~n}ez}, {Sagrist{\`a} Sell{\'e}s}, {Sahlmann}, {Salguero}, {Samaras},
  {Sanchez Gimenez}, {Sanna}, {Santove{\~n}a}, {Sarasso}, {Schultheis},
  {Sciacca}, {Segol}, {Segovia}, {S{\'e}gransan}, {Semeux}, {Shahaf},
  {Siddiqui}, {Siebert}, {Siltala}, {Silvelo}, {Slezak}, {Slezak}, {Snaith},
  {Solano}, {Solitro}, {Souami}, {Souchay}, {Spagna}, {Spina}, {Spoto},
  {Steele}, {Steidelm{\"u}ller}, {Stephenson}, {S{\"u}veges}, {Surdej},
  {Szabados}, {Szegedi-Elek}, {Taris}, {Taylor}, {Teixeira}, {Tolomei},
  {Tonello}, {Torra}, {Torra}, {Torralba Elipe}, {Trabucchi}, {Tsounis},
  {Turon}, {Ulla}, {Unger}, {Vaillant}, {van Dillen}, {van Reeven}, {Vanel},
  {Vecchiato}, {Viala}, {Vicente}, {Voutsinas}, {Weiler}, {Wevers},
  {Wyrzykowski}, {Yoldas}, {Yvard}, {Zhao}, {Zorec}, {Zucker}, \&
  {Zwitter}}]{2022arXiv220605870G}
{Gaia Collaboration}, {Creevey}, O.~L., {Sarro}, L.~M., {et~al.}
  2022{\natexlab{a}}, arXiv e-prints, arXiv:2206.05870

\bibitem[{{Gaia Collaboration} {et~al.}(2022{\natexlab{b}}){Gaia
  Collaboration}, {Vallenari}, {Brown}, {Prusti}, {de Bruijne}, {Arenou},
  {Babusiaux}, {Biermann}, {Creevey}, {Ducourant}, {Evans}, {Eyer}, {Guerra},
  {Hutton}, {Jordi}, {Klioner}, {Lammers}, {Lindegren}, {Luri}, {Mignard},
  {Panem}, {Pourbaix}, {Randich}, {Sartoretti}, {Soubiran}, {Tanga}, {Walton},
  {Bailer-Jones}, {Bastian}, {Drimmel}, {Jansen}, {Katz}, {Lattanzi}, {van
  Leeuwen}, {Bakker}, {Cacciari}, {Casta{\~n}eda}, {De Angeli}, {Fabricius},
  {Fouesneau}, {Fr{\'e}mat}, {Galluccio}, {Guerrier}, {Heiter}, {Masana},
  {Messineo}, {Mowlavi}, {Nicolas}, {Nienartowicz}, {Pailler}, {Panuzzo},
  {Riclet}, {Roux}, {Seabroke}, {Sordo{\o}rcit}, {Th{\'e}venin},
  {Gracia-Abril}, {Portell}, {Teyssier}, {Altmann}, {Andrae}, {Audard},
  {Bellas-Velidis}, {Benson}, {Berthier}, {Blomme}, {Burgess}, {Busonero},
  {Busso}, {C{\'a}novas}, {Carry}, {Cellino}, {Cheek}, {Clementini},
  {Damerdji}, {Davidson}, {de Teodoro}, {Nu{\~n}ez Campos}, {Delchambre},
  {Dell'Oro}, {Esquej}, {Fern{\'a}ndez-Hern{\'a}ndez}, {Fraile}, {Garabato},
  {Garc{\'\i}a-Lario}, {Gosset}, {Haigron}, {Halbwachs}, {Hambly}, {Harrison},
  {Hern{\'a}ndez}, {Hestroffer}, {Hodgkin}, {Holl}, {Jan{\ss}en}, {Jevardat de
  Fombelle}, {Jordan}, {Krone-Martins}, {Lanzafame}, {L{\"o}ffler}, {Marchal},
  {Marrese}, {Moitinho}, {Muinonen}, {Osborne}, {Pancino}, {Pauwels},
  {Recio-Blanco}, {Reyl{\'e}}, {Riello}, {Rimoldini}, {Roegiers}, {Rybizki},
  {Sarro}, {Siopis}, {Smith}, {Sozzetti}, {Utrilla}, {van Leeuwen}, {Abbas},
  {{\'A}brah{\'a}m}, {Abreu Aramburu}, {Aerts}, {Aguado}, {Ajaj},
  {Aldea-Montero}, {Altavilla}, {{\'A}lvarez}, {Alves}, {Anders}, {Anderson},
  {Anglada Varela}, {Antoja}, {Baines}, {Baker}, {Balaguer-N{\'u}{\~n}ez},
  {Balbinot}, {Balog}, {Barache}, {Barbato}, {Barros}, {Barstow},
  {Bartolom{\'e}}, {Bassilana}, {Bauchet}, {Becciani}, {Bellazzini},
  {Berihuete}, {Bernet}, {Bertone}, {Bianchi}, {Binnenfeld}, {Blanco-Cuaresma},
  {Blazere}, {Boch}, {Bombrun}, {Bossini}, {Bouquillon}, {Bragaglia},
  {Bramante}, {Breedt}, {Bressan}, {Brouillet}, {Brugaletta}, {Bucciarelli},
  {Burlacu}, {Butkevich}, {Buzzi}, {Caffau}, {Cancelliere}, {Cantat-Gaudin},
  {Carballo}, {Carlucci}, {Carnerero}, {Carrasco}, {Casamiquela}, {Castellani},
  {Castro-Ginard}, {Chaoul}, {Charlot}, {Chemin}, {Chiaramida}, {Chiavassa},
  {Chornay}, {Comoretto}, {Contursi}, {Cooper}, {Cornez}, {Cowell}, {Crifo},
  {Cropper}, {Crosta}, {Crowley}, {Dafonte}, {Dapergolas}, {David}, {David},
  {de Laverny}, {De Luise}, {De March}, {De Ridder}, {de Souza}, {de Torres},
  {del Peloso}, {del Pozo}, {Delbo}, {Delgado}, {Delisle}, {Demouchy},
  {Dharmawardena}, {Di Matteo}, {Diakite}, {Diener}, {Distefano}, {Dolding},
  {Edvardsson}, {Enke}, {Fabre}, {Fabrizio}, {Faigler}, {Fedorets}, {Fernique},
  {Fienga}, {Figueras}, {Fournier}, {Fouron}, {Fragkoudi}, {Gai},
  {Garcia-Gutierrez}, {Garcia-Reinaldos}, {Garc{\'\i}a-Torres}, {Garofalo},
  {Gavel}, {Gavras}, {Gerlach}, {Geyer}, {Giacobbe}, {Gilmore}, {Girona},
  {Giuffrida}, {Gomel}, {Gomez}, {Gonz{\'a}lez-N{\'u}{\~n}ez},
  {Gonz{\'a}lez-Santamar{\'\i}a}, {Gonz{\'a}lez-Vidal}, {Granvik}, {Guillout},
  {Guiraud}, {Guti{\'e}rrez-S{\'a}nchez}, {Guy}, {Hatzidimitriou}, {Hauser},
  {Haywood}, {Helmer}, {Helmi}, {Sarmiento}, {Hidalgo}, {Hilger},
  {H{\l}adczuk}, {Hobbs}, {Holland}, {Huckle}, {Jardine}, {Jasniewicz},
  {Jean-Antoine Piccolo}, {Jim{\'e}nez-Arranz}, {Jorissen}, {Juaristi
  Campillo}, {Julbe}, {Karbevska}, {Kervella}, {Khanna}, {Kontizas},
  {Kordopatis}, {Korn}, {K{\'o}sp{\'a}l}, {Kostrzewa-Rutkowska},
  {Kruszy{\'n}ska}, {Kun}, {Laizeau}, {Lambert}, {Lanza}, {Lasne}, {Le
  Campion}, {Lebreton}, {Lebzelter}, {Leccia}, {Leclerc}, {Lecoeur-Taibi},
  {Liao}, {Licata}, {Lindstr{\o}m}, {Lister}, {Livanou}, {Lobel}, {Lorca},
  {Loup}, {Madrero Pardo}, {Magdaleno Romeo}, {Managau}, {Mann}, {Manteiga},
  {Marchant}, {Marconi}, {Marcos}, {Marcos Santos}, {Mar{\'\i}n Pina},
  {Marinoni}, {Marocco}, {Marshall}, {Polo}, {Mart{\'\i}n-Fleitas}, {Marton},
  {Mary}, {Masip}, {Massari}, {Mastrobuono-Battisti}, {Mazeh}, {McMillan},
  {Messina}, {Michalik}, {Millar}, {Mints}, {Molina}, {Molinaro}, {Moln{\'a}r},
  {Monari}, {Mongui{\'o}}, {Montegriffo}, {Montero}, {Mor}, {Mora},
  {Morbidelli}, {Morel}, {Morris}, {Muraveva}, {Murphy}, {Musella}, {Nagy},
  {Noval}, {Oca{\~n}a}, {Ogden}, {Ordenovic}, {Osinde}, {Pagani}, {Pagano},
  {Palaversa}, {Palicio}, {Pallas-Quintela}, {Panahi}, {Payne-Wardenaar},
  {Pe{\~n}alosa Esteller}, {Penttil{\"a}}, {Pichon}, {Piersimoni}, {Pineau},
  {Plachy}, {Plum}, {Poggio}, {Pr{\v{s}}a}, {Pulone}, {Racero}, {Ragaini},
  {Rainer}, {Raiteri}, {Rambaux}, {Ramos}, {Ramos-Lerate}, {Re Fiorentin},
  {Regibo}, {Richards}, {Rios Diaz}, {Ripepi}, {Riva}, {Rix}, {Rixon},
  {Robichon}, {Robin}, {Robin}, {Roelens}, {Rogues}, {Rohrbasser},
  {Romero-G{\'o}mez}, {Rowell}, {Royer}, {Ruz Mieres}, {Rybicki}, {Sadowski},
  {S{\'a}ez N{\'u}{\~n}ez}, {Sagrist{\`a} Sell{\'e}s}, {Sahlmann}, {Salguero},
  {Samaras}, {Sanchez Gimenez}, {Sanna}, {Santove{\~n}a}, {Sarasso},
  {Schultheis}, {Sciacca}, {Segol}, {Segovia}, {S{\'e}gransan}, {Semeux},
  {Shahaf}, {Siddiqui}, {Siebert}, {Siltala}, {Silvelo}, {Slezak}, {Slezak},
  {Smart}, {Snaith}, {Solano}, {Solitro}, {Souami}, {Souchay}, {Spagna},
  {Spina}, {Spoto}, {Steele}, {Steidelm{\"u}ller}, {Stephenson}, {S{\"u}veges},
  {Surdej}, {Szabados}, {Szegedi-Elek}, {Taris}, {Taylo}, {Teixeira},
  {Tolomei}, {Tonello}, {Torra}, {Torra}, {Torralba Elipe}, {Trabucchi},
  {Tsounis}, {Turon}, {Ulla}, {Unger}, {Vaillant}, {van Dillen}, {van Reeven},
  {Vanel}, {Vecchiato}, {Viala}, {Vicente}, {Voutsinas}, {Weiler}, {Wevers},
  {Wyrzykowski}, {Yoldas}, {Yvard}, {Zhao}, {Zorec}, {Zucker}, \&
  {Zwitter}}]{GaiaDR3}
{Gaia Collaboration}, {Vallenari}, A., {Brown}, A.~G.~A., {et~al.}
  2022{\natexlab{b}}, arXiv e-prints, arXiv:2208.00211

\bibitem[{{Gaia Collaboration} {et~al.}(2022{\natexlab{c}}){Gaia
  Collaboration}, {Montegriffo}, {Bellazzini}, {De Angeli}, {Andrae},
  {Barstow}, {Bossini}, {Bragaglia}, {Burgess}, {Cacciari}, {Carrasco},
  {Chornay}, {Delchambre}, {Evans}, {Fouesneau}, {Fremat}, {Garabato}, {Jordi},
  {Manteiga}, {Massari}, {Palaversa}, {Pancino}, {Riello}, {Ruz Mieres},
  {Sanna}, {Santovena}, {Sordo}, {Vallenari}, {Walton}, \&
  {DPAC}}]{2022arXiv220606215G}
{Gaia Collaboration}, {Montegriffo}, P., {Bellazzini}, M., {et~al.}
  2022{\natexlab{c}}, arXiv e-prints, arXiv:2206.06215

\bibitem[{{Gallazzi} {et~al.}(2005){Gallazzi}, {Charlot}, {Brinchmann},
  {White}, \& {Tremonti}}]{Gallazzi2005}
{Gallazzi}, A., {Charlot}, S., {Brinchmann}, J., {White}, S. D.~M., \&
  {Tremonti}, C.~A. 2005, MNRAS, 362, 41

\bibitem[{{Gilmore} {et~al.}(2012){Gilmore}, {Randich}, {Asplund}, {Binney},
  {Bonifacio}, {Drew}, {Feltzing}, {Ferguson}, {Jeffries}, {Micela}, \&
  et~al.}]{Gilmore2012}
{Gilmore}, G., {Randich}, S., {Asplund}, M., {et~al.} 2012, Msngr, 147, 25

\bibitem[{Harris {et~al.}(2020)Harris, Millman, van~der Walt, Gommers,
  Virtanen, Cournapeau, Wieser, Taylor, Berg, Smith, Kern, Picus, Hoyer, van
  Kerkwijk, Brett, Haldane, del R{\'{i}}o, Wiebe, Peterson,
  G{\'{e}}rard-Marchant, Sheppard, Reddy, Weckesser, Abbasi, Gohlke, \&
  Oliphant}]{Harris2020}
Harris, C.~R., Millman, K.~J., van~der Walt, S.~J., {et~al.} 2020, Nature, 585,
  357

\bibitem[{{Hayden} {et~al.}(2015){Hayden}, {Bovy}, {Holtzman}, {Nidever},
  {Bird}, {Weinberg}, {Andrews}, {Majewski}, {Allende Prieto}, {Anders},
  {Beers}, {Bizyaev}, {Chiappini}, {Cunha}, {Frinchaboy},
  {Garc{\'\i}a-Her{\'n}andez}, {Garc{\'\i}a P{\'e}rez}, {Girardi}, {Harding},
  {Hearty}, {Johnson}, {M{\'e}sz{\'a}ros}, {Minchev}, {O'Connell}, {Pan},
  {Robin}, {Schiavon}, {Schneider}, {Schultheis}, {Shetrone}, {Skrutskie},
  {Steinmetz}, {Smith}, {Wilson}, {Zamora}, \& {Zasowski}}]{Hayden2015}
{Hayden}, M.~R., {Bovy}, J., {Holtzman}, J.~A., {et~al.} 2015, ApJ, 808, 132

\bibitem[{Hunter(2007)}]{Hunter2007}
Hunter, J.~D. 2007, Computing in Science \& Engineering, 9, 90

\bibitem[{{Ji} {et~al.}(2019{\natexlab{a}}){Ji}, {Drout}, \&
  {Hansen}}]{Ji2019a}
{Ji}, A.~P., {Drout}, M.~R., \& {Hansen}, T.~T. 2019{\natexlab{a}}, ApJ, 882,
  40

\bibitem[{{Ji} {et~al.}(2019{\natexlab{b}}){Ji}, {Simon}, {Frebel}, {Venn}, \&
  {Hansen}}]{Ji2019b}
{Ji}, A.~P., {Simon}, J.~D., {Frebel}, A., {Venn}, K.~A., \& {Hansen}, T.~T.
  2019{\natexlab{b}}, ApJ, 870, 83

\bibitem[{{Katz} {et~al.}(2022){Katz}, {Sartoretti}, {Guerrier}, {Panuzzo},
  {Seabroke}, {Th{\'e}venin}, {Cropper}, {Benson}, {Blomme}, {Haigron},
  {Marchal}, {Smith}, {Baker}, {Chemin}, {Damerdji}, {David}, {Dolding},
  {Fr{\'e}mat}, {Gosset}, {Jan{\ss}en}, {Jasniewicz}, {Lobel}, {Plum},
  {Samaras}, {Snaith}, {Soubiran}, {Vanel}, {Zwitter}, {Antoja}, {Arenou},
  {Babusiaux}, {Brouillet}, {Caffau}, {Di Matteo}, {Fabre}, {Fabricius},
  {Frakgoudi}, {Haywood}, {Huckle}, {Hottier}, {Lasne}, {Leclerc},
  {Mastrobuono-Battisti}, {Royer}, {Teyssier}, {Zorec}, {Crifo}, {Jean-Antoine
  Piccolo}, {Turon}, \& {Viala}}]{2022arXiv220605902K}
{Katz}, D., {Sartoretti}, P., {Guerrier}, A., {et~al.} 2022, arXiv e-prints,
  arXiv:2206.05902

\bibitem[{{Kollmeier} {et~al.}(2017){Kollmeier}, {Zasowski}, {Rix}, {Johns},
  {Anderson}, {Drory}, {Johnson}, {Pogge}, {Bird}, {Blanc}, {Brownstein},
  {Crane}, {De Lee}, {Klaene}, {Kreckel}, {MacDonald}, {Merloni}, {Ness},
  {O'Brien}, {Sanchez-Gallego}, {Sayres}, {Shen}, {Thakar}, {Tkachenko},
  {Aerts}, {Blanton}, {Eisenstein}, {Holtzman}, {Maoz}, {Nandra}, {Rockosi},
  {Weinberg}, {Bovy}, {Casey}, {Chaname}, {Clerc}, {Conroy}, {Eracleous},
  {G{\"a}nsicke}, {Hekker}, {Horne}, {Kauffmann}, {McQuinn}, {Pellegrini},
  {Schinnerer}, {Schlafly}, {Schwope}, {Seibert}, {Teske}, \& {van
  Saders}}]{SDSSV}
{Kollmeier}, J.~A., {Zasowski}, G., {Rix}, H.-W., {et~al.} 2017, arXiv
  e-prints, arXiv:1711.03234

\bibitem[{{Li} {et~al.}(2022){Li}, {Aoki}, {Matsuno}, {Xing}, {Suda},
  {Tominaga}, {Chen}, {Honda}, {Ishigaki}, {Shi}, {Zhao}, \&
  {Zhao}}]{2022ApJ...931..147L}
{Li}, H., {Aoki}, W., {Matsuno}, T., {et~al.} 2022, \apj, 931, 147

\bibitem[{{Lindegren} {et~al.}(2021){Lindegren}, {Bastian}, {Biermann},
  {Bombrun}, {de Torres}, {Gerlach}, {Geyer}, {Hern{\'a}ndez}, {Hilger},
  {Hobbs}, {Klioner}, {Lammers}, {McMillan}, {Ramos-Lerate},
  {Steidelm{\"u}ller}, {Stephenson}, \& {van Leeuwen}}]{2021A&A...649A...4L}
{Lindegren}, L., {Bastian}, U., {Biermann}, M., {et~al.} 2021, \aap, 649, A4

\bibitem[{{Marocco} {et~al.}(2021){Marocco}, {Eisenhardt}, {Fowler},
  {Kirkpatrick}, {Meisner}, {Schlafly}, {Stanford}, {Garcia}, {Caselden},
  {Cushing}, {Cutri}, {Faherty}, {Gelino}, {Gonzalez}, {Jarrett}, {Koontz},
  {Mainzer}, {Marchese}, {Mobasher}, {Schlegel}, {Stern}, {Teplitz}, \&
  {Wright}}]{2021ApJS..253....8M}
{Marocco}, F., {Eisenhardt}, P. R.~M., {Fowler}, J.~W., {et~al.} 2021, \apjs,
  253, 8

\bibitem[{{Matteucci}(1994)}]{Matteucci1994}
{Matteucci}, F. 1994, A\&A, 288, 57

\bibitem[{{McWilliam}(1997)}]{McWilliam1997}
{McWilliam}, A. 1997, ARA\&A, 35, 503

\bibitem[{{Montegriffo} {et~al.}(2022){Montegriffo}, {De Angeli}, {Andrae},
  {Riello}, {Pancino}, {Sanna}, {Bellazzini}, {Evans}, {Carrasco}, {Sordo},
  {Busso}, {Cacciari}, {Jordi}, {van Leeuwen}, {Vallenari}, {Altavilla},
  {Barstow}, {Brown}, {Burgess}, {Castellani}, {Cowell}, {Davidson}, {De
  Luise}, {Delchambre}, {Diener}, {Fabricius}, {Fremat}, {Fouesneau},
  {Gilmore}, {Giuffrida}, {Hambly}, {Harrison}, {Hidalgo}, {Hodgkin},
  {Holland}, {Marinoni}, {Osborne}, {Pagani}, {Palaversa}, {Piersimoni},
  {Pulone}, {Ragaini}, {Rainer}, {Richards}, {Rowell}, {Ruz-Mieres}, {Sarro},
  {Walton}, \& {Yoldas}}]{Montegriffo2022}
{Montegriffo}, P., {De Angeli}, F., {Andrae}, R., {et~al.} 2022, arXiv
  e-prints, arXiv:2206.06205

\bibitem[{{Ness} {et~al.}(2019){Ness}, {Johnston}, {Blancato}, {Rix}, {Beane},
  {Bird}, \& {Hawkins}}]{Ness2019}
{Ness}, M.~K., {Johnston}, K.~V., {Blancato}, K., {et~al.} 2019, ApJ, 883, 177

\bibitem[{{Nordstr{\"o}m} {et~al.}(2004){Nordstr{\"o}m}, {Mayor}, {Andersen},
  {Holmberg}, {Pont}, {J{\o}rgensen}, {Olsen}, {Udry}, \&
  {Mowlavi}}]{Nordstrom2004}
{Nordstr{\"o}m}, B., {Mayor}, M., {Andersen}, J., {et~al.} 2004, A\&A, 418, 989

\bibitem[{{Recio-Blanco} {et~al.}(2022){Recio-Blanco}, {de Laverny}, {Palicio},
  {Kordopatis}, {{\'A}lvarez}, {Schultheis}, {Contursi}, {Zhao}, {Torralba
  Elipe}, {Ordenovic}, {Manteiga}, {Dafonte}, {Oreshina-Slezak}, {Bijaoui},
  {Fremat}, {Seabroke}, {Pailler}, {Spitoni}, {Poggio}, {Creevey}, {Abreu
  Aramburu}, {Accart}, {Andrae}, {Bailer-Jones}, {Bellas-Velidis}, {Brouillet},
  {Brugaletta}, {Burlacu}, {Carballo}, {Casamiquela}, {Chiavassa}, {Cooper},
  {Dapergolas}, {Delchambre}, {Dharmawardena}, {Drimmel}, {Edvardsson},
  {Fouesneau}, {Garabato}, {Garcia-Lario}, {Garcia-Torres}, {Gavel}, {Gomez},
  {Gonzalez-Santamaria}, {Hatzidimitriou}, {Heiter}, {Jean-Antoine Piccolo},
  {Kontizas}, {Korn}, {Lanzafame}, {Lebreton}, {Le Fustec}, {Licata},
  {Lindstrom}, {Livanou}, {Lobel}, {Lorca}, {Magdaleno Romeo}, {Marocco},
  {Marshall}, {Mary}, {Nicolas}, {Pallas-Quintela}, {Panem}, {Pichon},
  {Riclet}, {Robin}, {Rybizki}, {Santovena}, {Silvelo}, {Smart}, {Sarro},
  {Sordo}, {Soubiran}, {Suvege}, {Ulla}, {Vallenari}, {Zorec}, {Utrilla}, \&
  {Bakker}}]{Recio-Blanco-GSPspec}
{Recio-Blanco}, A., {de Laverny}, P., {Palicio}, P.~A., {et~al.} 2022, arXiv
  e-prints, arXiv:2206.05541

\bibitem[{{Rix} {et~al.}(2021){Rix}, {Hogg}, {Boubert}, {Brown}, {Casey},
  {Drimmel}, {Everall}, {Fouesneau}, \& {Price-Whelan}}]{Rix2021}
{Rix}, H.-W., {Hogg}, D.~W., {Boubert}, D., {et~al.} 2021, \aj, 162, 142

\bibitem[{{Rix} {et~al.}(2022){Rix}, {Chandra}, {Andrae}, {Price-Whelan},
  {Weinberg}, {Conroy}, {Fouesneau}, {Hogg}, {De Angeli}, {Naidu}, {Xiang}, \&
  {Ruz-Mieres}}]{Rix2022}
{Rix}, H.-W., {Chandra}, V., {Andrae}, R., {et~al.} 2022, \apj, 941, 45

\bibitem[{{Sartoretti} {et~al.}(2022){Sartoretti}, {Blomme}, {David}, \&
  {Seabroke}}]{2022gdr3.reptE...6S}
{Sartoretti}, P., {Blomme}, R., {David}, M., \& {Seabroke}, G. 2022, {Gaia DR3
  documentation Chapter 6: Spectroscopy}, ,

\bibitem[{{Sneden} {et~al.}(2008){Sneden}, {Cowan}, \& {Gallino}}]{Sneden2008}
{Sneden}, C., {Cowan}, J.~J., \& {Gallino}, R. 2008, ARA\&A, 46, 241

\bibitem[{{Ting} {et~al.}(2017){Ting}, {Conroy}, {Rix}, \&
  {Cargile}}]{Ting2017}
{Ting}, Y.-S., {Conroy}, C., {Rix}, H.-W., \& {Cargile}, P. 2017, \apj, 843, 32

\bibitem[{{Ting} \& {Weinberg}(2022)}]{Ting2022}
{Ting}, Y.-S., \& {Weinberg}, D.~H. 2022, ApJ, 927, 209

\bibitem[{{Tinsley}(1979)}]{Tinsley1979}
{Tinsley}, B.~M. 1979, ApJ, 229, 1046

\bibitem[{{Tinsley}(1980)}]{Tinsley1980}
---. 1980, FCPh, 5, 287

\bibitem[{{Tremonti} {et~al.}(2004){Tremonti}, {Heckman}, {Kauffmann},
  {Brinchmann}, {Charlot}, {White}, {Seibert}, {Peng}, {Schlegel}, {Uomoto},
  {Fukugita}, \& {Brinkmann}}]{Tremonti2004}
{Tremonti}, C.~A., {Heckman}, T.~M., {Kauffmann}, G., {et~al.} 2004, ApJ, 613,
  898

\bibitem[{{Twarog}(1980)}]{Twarog1980}
{Twarog}, B.~A. 1980, ApJ, 242, 242

\bibitem[{Virtanen {et~al.}(2020)Virtanen, Gommers, Oliphant, Haberland, Reddy,
  Cournapeau, Burovski, Peterson, Weckesser, Bright, van~der Walt, Brett,
  Wilson, Millman, Mayorov, Nelson, Jones, Kern, Larson, Carey, Polat, Feng,
  Moore, VanderPlas, Laxalde, Perktold, Cimrman, Henriksen, Quintero, Harris,
  Archibald, Ribeiro, Pedregosa, van Mulbregt, Vijaykumar, Bardelli, Rothberg,
  Hilboll, Kloeckner, Scopatz, Lee, Rokem, Woods, Fulton, Masson,
  H{\"{a}}ggstr{\"{o}}m, Fitzgerald, Nicholson, Hagen, Pasechnik, Olivetti,
  Martin, Wieser, Silva, Lenders, Wilhelm, Young, Price, Ingold, Allen, Lee,
  Audren, Probst, Dietrich, Silterra, Webber, Slavi{\v{c}}, Nothman, Buchner,
  Kulick, Sch{\"{o}}nberger, {de Miranda Cardoso}, Reimer, Harrington,
  Rodr{\'{i}}guez, Nunez-Iglesias, Kuczynski, Tritz, Thoma, Newville,
  K{\"{u}}mmerer, Bolingbroke, Tartre, Pak, Smith, Nowaczyk, Shebanov, Pavlyk,
  Brodtkorb, Lee, McGibbon, Feldbauer, Lewis, Tygier, Sievert, Vigna, Peterson,
  More, Pudlik, Oshima, Pingel, Robitaille, Spura, Jones, Cera, Leslie, Zito,
  Krauss, Upadhyay, Halchenko, \& V{\'{a}}zquez-Baeza}]{Virtanen2020}
Virtanen, P., Gommers, R., Oliphant, T.~E., {et~al.} 2020, Nature Methods, 17,
  261

\bibitem[{{Weinberg} {et~al.}(2017){Weinberg}, {Andrews}, \&
  {Freudenburg}}]{Weinberg2017}
{Weinberg}, D.~H., {Andrews}, B.~H., \& {Freudenburg}, J. 2017, \apj, 837, 183

\bibitem[{{Weinberg} {et~al.}(2019){Weinberg}, {Holtzman}, {Hasselquist},
  {Bird}, {Johnson}, {Shetrone}, {Sobeck}, {Allende Prieto}, {Bizyaev},
  {Carrera}, {Cohen}, {Cunha}, {Ebelke}, {Fernandez-Trincado},
  {Garc{\'\i}a-Hern{\'a}ndez}, {Hayes}, {J{\"o}nsson}, {Lane}, {Majewski},
  {Malanushenko}, {M{\'e}sz{\'a}ros}, {Nidever}, {Nitschelm}, {Pan}, {Rix},
  {Rybizki}, {Schiavon}, {Schneider}, {Wilson}, \& {Zamora}}]{Weinberg2019}
{Weinberg}, D.~H., {Holtzman}, J.~A., {Hasselquist}, S., {et~al.} 2019, ApJ,
  874, 102

\bibitem[{{Weinberg} {et~al.}(2022){Weinberg}, {Holtzman}, {Johnson}, {Hayes},
  {Hasselquist}, {Shetrone}, {Ting}, {Beaton}, {Beers}, {Bird}, {Bizyaev},
  {Blanton}, {Cunha}, {Fern{\'a}ndez-Trincado}, {Frinchaboy},
  {Garc{\'\i}a-Hern{\'a}ndez}, {Griffith}, {Johnson}, {J{\"o}nsson}, {Lane},
  {Leung}, {Mackereth}, {Majewski}, {M{\'e}sz{\'a}ros}, {Nitschelm}, {Pan},
  {Schiavon}, {Schneider}, {Schultheis}, {Smith}, {Sobeck}, {Stassun},
  {Stringfellow}, {Vincenzo}, {Wilson}, \& {Zasowski}}]{Weinberg2022}
{Weinberg}, D.~H., {Holtzman}, J.~A., {Johnson}, J.~A., {et~al.} 2022, ApJS,
  260, 32

\bibitem[{{Wu} {et~al.}(2014){Wu}, {Du}, {Luo}, {Zhao}, \&
  {Yuan}}]{2014IAUS..306..340W}
{Wu}, Y., {Du}, B., {Luo}, A., {Zhao}, Y., \& {Yuan}, H. 2014, in IAU
  Symposium, Vol. 306, Statistical Challenges in 21st Century Cosmology, ed.
  A.~{Heavens}, J.-L. {Starck}, \& A.~{Krone-Martins}, 340--342

\bibitem[{{Wu} {et~al.}(2011){Wu}, {Luo}, {Li}, {Shi}, {Prugniel}, {Liang},
  {Zhao}, {Zhang}, {Bai}, {Wei}, {Dong}, {Zhang}, \&
  {Chen}}]{2011RAA....11..924W}
{Wu}, Y., {Luo}, A.-L., {Li}, H.-N., {et~al.} 2011, Research in Astronomy and
  Astrophysics, 11, 924

\bibitem[{{Yao} {et~al.}(2023){Yao}, {Ji}, {Koposov}, \&
  {Limberg}}]{2023arXiv230317676Y}
{Yao}, Y., {Ji}, A.~P., {Koposov}, S.~E., \& {Limberg}, G. 2023, arXiv
  e-prints, arXiv:2303.17676

\bibitem[{{York} {et~al.}(2000){York}, {Adelman}, {Anderson}, {Anderson},
  {Annis}, {Bahcall}, {Bakken}, {Barkhouser}, {Bastian}, {Berman}, {Boroski},
  {Bracker}, {Briegel}, {Briggs}, {Brinkmann}, {Brunner}, {Burles}, {Carey},
  {Carr}, {Castander}, {Chen}, {Colestock}, {Connolly}, {Crocker}, {Csabai},
  {Czarapata}, {Davis}, {Doi}, {Dombeck}, {Eisenstein}, {Ellman}, {Elms},
  {Evans}, {Fan}, {Federwitz}, {Fiscelli}, {Friedman}, {Frieman}, {Fukugita},
  {Gillespie}, {Gunn}, {Gurbani}, {de Haas}, {Haldeman}, {Harris}, {Hayes},
  {Heckman}, {Hennessy}, {Hindsley}, {Holm}, {Holmgren}, {Huang}, {Hull},
  {Husby}, {Ichikawa}, {Ichikawa}, {Ivezi{\'c}}, {Kent}, {Kim}, {Kinney},
  {Klaene}, {Kleinman}, {Kleinman}, {Knapp}, {Korienek}, {Kron}, {Kunszt},
  {Lamb}, {Lee}, {Leger}, {Limmongkol}, {Lindenmeyer}, {Long}, {Loomis},
  {Loveday}, {Lucinio}, {Lupton}, {MacKinnon}, {Mannery}, {Mantsch}, {Margon},
  {McGehee}, {McKay}, {Meiksin}, {Merelli}, {Monet}, {Munn}, {Narayanan},
  {Nash}, {Neilsen}, {Neswold}, {Newberg}, {Nichol}, {Nicinski}, {Nonino},
  {Okada}, {Okamura}, {Ostriker}, {Owen}, {Pauls}, {Peoples}, {Peterson},
  {Petravick}, {Pier}, {Pope}, {Pordes}, {Prosapio}, {Rechenmacher}, {Quinn},
  {Richards}, {Richmond}, {Rivetta}, {Rockosi}, {Ruthmansdorfer}, {Sandford},
  {Schlegel}, {Schneider}, {Sekiguchi}, {Sergey}, {Shimasaku}, {Siegmund},
  {Smee}, {Smith}, {Snedden}, {Stone}, {Stoughton}, {Strauss}, {Stubbs},
  {SubbaRao}, {Szalay}, {Szapudi}, {Szokoly}, {Thakar}, {Tremonti}, {Tucker},
  {Uomoto}, {Vanden Berk}, {Vogeley}, {Waddell}, {Wang}, {Watanabe},
  {Weinberg}, {Yanny}, {Yasuda}, \& {SDSS Collaboration}}]{York2000}
{York}, D.~G., {Adelman}, J., {Anderson}, John~E., J., {et~al.} 2000, AJ, 120,
  1579

\bibitem[{{Zhang} {et~al.}(2023){Zhang}, {Green}, \&
  {Rix}}]{2023arXiv230303420Z}
{Zhang}, X., {Green}, G.~M., \& {Rix}, H.-W. 2023, arXiv e-prints,
  arXiv:2303.03420

\end{thebibliography}
\bibliographystyle{aasjournal}

\appendix

\section{Details of XGBoost}
\label{appendix:details-XGBoost}

In order to ensure reproducibility, this appendix provides further details about our XGBoost model. First and foremost, we list the input features used for all XGBoost models (\MH, \Teff, \logg):
\begin{itemize}
\item 110 XP coefficients, each divided by $10^{(15 - G)/2.5}$ for normalisation
\item 5 features of the form of Eq.~(\ref{eq:feature-with-parallax}) for $G$, $G_\textrm{BP}$, $G_\textrm{RP}$, $W_1$ and $W_2$
\item 7 observed colours: $G-G_\textrm{BP}$
    $G - G_\textrm{RP}$,
    $G_\textrm{BP} - G_\textrm{RP}$,
    $G_\textrm{BP} - W_2$,
    $G - W_1$,
    $G - W_2$,
    $W_1   - W_2$
\item 31 synthesised colours, each of the form of $G-X$, for (using GaiaXPy nomenclature) StromgrenStd\_mag\_b,
StromgrenStd\_mag\_y,
Jplus\_mag\_gJPLUS,
Jplus\_mag\_iJPLUS,
Jplus\_mag\_J0515,
Jplus\_mag\_J0861,
Jplus\_mag\_J0660,
Panstarrs1Std\_mag\_gp,
Panstarrs1Std\_mag\_rp,
Panstarrs1Std\_mag\_ip,
Panstarrs1Std\_mag\_zp,
Panstarrs1Std\_mag\_yp,
SkyMapper\_mag\_g,
SkyMapper\_mag\_r,
SkyMapper\_mag\_i,
SkyMapper\_mag\_z,
Gaia2\_mag\_C1B431,
Gaia2\_mag\_C1B556,
Gaia2\_mag\_C1B655,
Gaia2\_mag\_C1B768,
Gaia2\_mag\_C1B916,
Gaia2\_mag\_C1M467,
Gaia2\_mag\_C1M506,
Gaia2\_mag\_C1M515,
Gaia2\_mag\_C1M549,
Gaia2\_mag\_C1M656,
Gaia2\_mag\_C1M716,
Gaia2\_mag\_C1M747,
Gaia2\_mag\_C1M825,
Gaia2\_mag\_C1M861,
Gaia2\_mag\_C1M965
\end{itemize}
Concerning the details of the XGBoost models themselves, the following Python code shows how the training was configured:
\begin{verbatim}
from sklearn.experimental
    import enable_hist_gradient_boosting 
from sklearn.ensemble 
    import HistGradientBoostingRegressor

xgboost = HistGradientBoostingRegressor(
             loss='least_squares',
             min_samples_leaf=20,
             max_depth=50, 
             max_leaf_nodes=500,
             max_iter=10000, 
             max_bins=255,
             l2_regularization=1.0e-9
          )
\end{verbatim}
Specifically for training the \MH\ model, we weighted the training examples with $e^{-\MH/5}$ in order to put more emphasis on low-metallicity examples:
\begin{verbatim}
Weights = numpy.exp(-MH/5)
modelMH = xgboost.fit(Features, MH,
          sample_weight=Weights)
\end{verbatim}
The XGBoost models for \Teff\ and \logg\ were configured in exactly the same way but did not use any weighting during training.

\section{Extracting Gaia DR3 sources with radial velocities and astrometry}
\label{appendix:ADQL}

For chemodynamical studies, the reader may want to match our \MH\ estimates to stars that have radial velocity measurements and astrometry in Gaia DR3. Table~\ref{table:vetted-RGB-results} contains all the information necessary to calculate orbits (for a given potential), but ``only'' for 13.3 million vetted RGB stars. Our full Table~\ref{table:full-results} contains only source IDs and  stellar parameter estimates (in particular \MH). Due to user quota limitations, it is impossible to upload either Table~\ref{table:vetted-RGB-results} or Table~\ref{table:full-results} to the Gaia archive for direct cross-matching. Instead, the user needs to download the data from the Gaia archive\footnote{\url{https://gea.esac.esa.int/archive/}} and then match via source\_id locally. The following ADQL query retrieves astrometry and radial velocity for 33\,653\,049 stars that have both available in Gaia DR3:
\begin{verbatim}
SELECT
source_id,
ra,dec,
parallax,parallax_error,
pmra,pmra_error,
pmdec,pmdec_error,
radial_velocity,radial_velocity_error
FROM gaiadr3.gaia_source_lite
WHERE radial_velocity IS NOT NULL 
    AND parallax IS NOT NULL
ORDER BY source_id
\end{verbatim}
The last line ensures that the results are sorted by Gaia source\_id in ascending order. This greatly facilitates the local cross-match for the reader, e.g.\ allowing for binary search during cross-matching.

\end{document}